\documentclass[10pt,conference]{IEEEtran}

\newcommand{\longv}[1]{}
\newcommand{\algo}[1]{}

\usepackage{flushend}
\usepackage{xspace}
\usepackage{graphicx}
\usepackage{color}
\usepackage[colorlinks=true,urlcolor=black,linkcolor=black,citecolor=black]{hyperref}
\usepackage{booktabs}
\usepackage{xspace}
\usepackage{graphicx}
\usepackage[caption=false]{subfig}
\usepackage{balance}
\usepackage{microtype}
\usepackage[inline]{enumitem}
\usepackage{listings,multicol}
\usepackage{xcolor}
\usepackage{url}
\usepackage{relsize}
\usepackage{adjustbox}
\usepackage{array}
\usepackage{csquotes}
\usepackage{threeparttable}
\usepackage{chngpage}
\usepackage{tabularx}
\usepackage{multirow}
\usepackage{algorithm,algorithmicx,algpseudocode}
\usepackage{chngpage}
\usepackage{comment}
\usepackage{amsmath}
\usepackage{amsfonts}
\usepackage{array}
\usepackage[T1]{fontenc}
\usepackage{ragged2e}
\usepackage[symbol]{footmisc}

\newcolumntype{P}[1]{>{\RaggedRight\hspace{0pt}}p{#1}}

\newcommand{\secref}[1]{Sec.\,\ref{#1}}

\newcommand{\figref}[1]{Fig.\,\ref{#1}}

\newcommand{\tabref}[1]{Table\,\ref{#1}}
\newcommand{\Figref}[1]{Figure\,\ref{#1}}

\newcommand{\edit}[1]{{\color{black} #1}}

\newcommand\parhead[1]{\vspace{.26mm}\noindent\textbf{{#1}}}  
\newcommand\myparhead[1]{\smallskip\noindent\textbf{{#1}}}  

\newcommand{\rl}[1]{{\textsf{\textbf{RL}}[\smaller\sffamily\color{blue} #1}]}
\newcommand{\tb}[1]{{\textsf{\textbf{TB}}[\smaller\sffamily\color{blue} #1}]}
\newcommand{\wm}[1]{{\textsf{\textbf{WM}}[\smaller\sffamily\color{red} #1}]}
\newcommand{\ds}[1]{{\textsf{\textbf{DS}}[\smaller\sffamily\color{violet} #1}]}

\def\inline{\lstinline[basicstyle=\ttfamily\fontsize{8}{8}\selectfont,,keywordstyle={}]}
\newcommand{\hashifdef}{\inline{\#ifdef}\xspace}

\newcommand{\co}{clone\&own\xspace}
\newcommand{\Co}{Clone\&own\xspace}

\newcommand{\bool}{\mathbb{B}}

\newcommand{\asset}{\textsl{asset}\xspace}
\newcommand{\featurecmd}{\textsl{feature}\xspace}
\newcommand{\featuremodel}{\textsl{feature model}\xspace}
\newcommand{\assettype}{\textsl{AssetType}\xspace}
\newcommand{\name}{\textsl{name}\xspace}
\newcommand{\version}{\textsl{version}\xspace}
\newcommand{\vproot}{\textsl{VPRootType}\xspace}
\newcommand{\repository}{\textsl{RepositoryType}\xspace}
\newcommand{\folder}{\textsl{FolderType}\xspace}
\newcommand{\file}{\textsl{FileType}\xspace}
\newcommand{\class}{\textsl{ClassType}\xspace}
\newcommand{\method}{\textsl{MethodType}\xspace}
\newcommand{\block}{\textsl{BlockType}\xspace}
\newcommand{\tracedb}{\textsl{AssetTraceDatabase}\xspace}
\newcommand{\ftracedb}{\textsl{FeatureTraceDatabase}\xspace}
\newcommand{\assettraces}{\textsl{AssetTraces}\xspace}

\newcommand{\presencecondition}{\textsl{presencecondition}\xspace}
\newcommand{\subassets}{\textsl{sub-assets}\xspace}
\newcommand{\subfeatures}{\textsl{sub-features}\xspace}
\newcommand{\globalversion}{\textsl{GlobalVersion}\xspace}
\newcommand{\parent}{\textsl{parent}\xspace}
\newcommand{\optional}{\textsl{optional}\xspace}
\newcommand{\rootnode}{\textit{root}\xspace}   

\newcommand{\incomplete}{\textsl{incomplete}\xspace}
\newcommand{\mappedfeatures}{\textsl{mappedfeatures}\xspace}
\newcommand{\mappedassets}{\textsl{mappedassets}\xspace}

\newcommand{\containable}{\textsl{containable}}
\newcommand{\wellformedness}{\textsl{well-formedness}}
\newcommand{\updateversion}{\textsl{updateversion}}
\newcommand{\add}{\textsl{add}}
\newcommand{\detectchanges}{\textsl{detectChanges}}

\newcommand\Tstrut{\rule{0pt}{2.6ex}}

\newcommand{\vpoperator}[1]{\texttt{#1}}
\newcommand{\addasset}{\vpoperator{AddAsset}}
\newcommand{\changeasset}{\vpoperator{ChangeAsset}\xspace}
\newcommand{\removeasset}{\vpoperator{RemoveAsset}\xspace}
\newcommand{\moveasset}{\vpoperator{MoveAsset}\xspace}
\newcommand{\addfeaturemodeltoasset}{\vpoperator{AddFeature\-Model\-To\-Asset}\xspace}
\newcommand{\mapassettofeature}{\vpoperator{MapAssetToFeature}\xspace}
\newcommand{\cloneasset}{\vpoperator{CloneAsset}\xspace}
\newcommand{\propagateasset}{\vpoperator{PropagateToAsset}\xspace}

\newcommand{\addfeature}{\vpoperator{AddFeature}\xspace}
\newcommand{\removefeature}{\vpoperator{RemoveFeature}\xspace}
\newcommand{\movefeature}{\vpoperator{MoveFeature}\xspace}
\newcommand{\makefeatureoptional}{\vpoperator{MakeFeatureOptional}\xspace}
\newcommand{\clonefeature}{\vpoperator{CloneFeature}\xspace}
\newcommand{\propagatefeature}{\vpoperator{PropagateToFeature}\xspace}

\newcommand{\getAncestorFeatureModel}{\vpoperator{getAncestorFeatureModel}\xspace}
\newcommand{\addunassignedfeature}{\vpoperator{addUnassignedFeature}\xspace}
\newcommand{\addtrace}{\vpoperator{addTrace}\xspace}
\newcommand{\featureexists}{\vpoperator{featureExists}\xspace}
\newcommand{\featurecloned}{\vpoperator{featureCloned}\xspace}
\newcommand{\assetcloned}{\vpoperator{assetCloned}\xspace}
\newcommand{\getclone}{\vpoperator{getClone}\xspace}
\newcommand{\getfeature}{\vpoperator{getFeature}\xspace}
\newcommand{\getlatesttrace}{\vpoperator{getLatestTrace}\xspace}
\newcommand{\makeconsistent}{\vpoperator{makeConsistent}\xspace}
\newcommand{\unmapasset}{\vpoperator{unmapAsset}\xspace}
\newcommand{\deletefeature}{\vpoperator{deleteFeature}\xspace}
\newcommand{\getslice}{\vpoperator{getSlice}\xspace}
\newcommand{\getfeaturepath}{\vpoperator{getFeaturePath}\xspace}
\newcommand{\isclone}{\vpoperator{isClone}\xspace}
\newcommand{\traversepath}{\vpoperator{traversePath}\xspace}

\newcommand{\Unassigned}{\textsl{Unassigned}}
\newcommand{\getmappedassets}{\vpoperator{getMappedAssets}\xspace}

\title{Seamless Variability Management\\With the Virtual Platform}

\author{\IEEEauthorblockN{Wardah Mahmood\IEEEauthorrefmark{1}, Daniel Strüber\IEEEauthorrefmark{2}, Thorsten Berger\IEEEauthorrefmark{1}\IEEEauthorrefmark{3}, Ralf Lämmel\IEEEauthorrefmark{4}, and Mukelabai Mukelabai\IEEEauthorrefmark{1}}
	\IEEEauthorblockA{\IEEEauthorrefmark{1}Chalmers\,|\,University of Gothenburg, Sweden}
	\IEEEauthorblockA{\IEEEauthorrefmark{2}Radboud University, Netherlands}
	\IEEEauthorblockA{\IEEEauthorrefmark{3}Ruhr University Bochum, Germany}
	\IEEEauthorblockA{\IEEEauthorrefmark{4}University of Koblenz-Landau, Germany}
}

\usepackage[absolute,overlay]{textpos}

\begin{document}
\begin{textblock*}{20cm}(0.9cm,0.5cm) 
   \textit{Author preprint for a paper published in the proceedings of the 43rd International Conference on Software Engineering (ICSE 2021)}
\end{textblock*}

\maketitle

\begin{abstract}
\looseness=-1
Customization is a general trend in software engineering, demanding systems that support variable stakeholder requirements. Two opposing strategies are commonly used to create variants: software \co and software configuration with an integrated platform. Organizations often start with the former, which is cheap, agile, and supports quick innovation, but does not scale. The latter scales by establishing an integrated platform that shares software assets between variants, but requires high up-front investments or risky migration processes. So, could we have a method that allows an easy transition or even combine the benefits of both strategies?
We propose a method and tool that supports a truly incremental development of variant-rich systems, exploiting a spectrum between
both opposing strategies. We design, formalize, and prototype the variability-management framework \textit{virtual platform}. It bridges \co and platform-oriented development.
Relying on programming-language-independent conceptual structures representing software assets, it offers operators for engineering and evolving a system, comprising: traditional, asset-oriented operators and novel, feature-oriented operators for incrementally adopting concepts of an integrated platform.
The operators record meta-data that is exploited by other operators to support the transition. Among others, they eliminate expensive feature-location effort or the need to trace clones.
Our evaluation simulates the evolution of a real-world, clone-based system, measuring its costs and benefits.
\end{abstract}


\section{Introduction}
\label{sec:introduction}
\noindent
\looseness=-1
Software systems often need to exist in many different variants. Organizations create variants to adapt systems to varying stakeholder requirements---for instance, to address a variety of market segments, runtime environments or different hardware. Creating variants allows organizations to experiment with new ideas and to test them on the market, which easily leads to a portfolio of system variants that needs to be maintained. 

\looseness=-1
Two opposing strategies exist for engineering variants. A convenient and frequent strategy is \emph{\co}\,\cite{dubinski.ea:2013:cloning,stuanciulescu2015forked,businge.ea:2018:appfamilies,jacob2020costs,lodewijksanalysis}, where developers create one system and then clone and adapt it to the new requirements. This strategy is well-supported by current version-control systems and tools, such as GIT, relying on their forking, branching, merging, and pull request facilities.
The frequent adoption of \co\,\cite{berger2013survey,dubinski.ea:2013:cloning,jacob2020costs} is usually attributed to its inexpensiveness, flexibility, and provided developer independence.
However, \co does not scale with the number of variants and then imposes substantial maintenance overheads.
\longv{Lacking any kind of centralized control, simple tasks such as bug fixing, refactoring or propagating new functionality (which can be scattered across the codebase) to other variants becomes redundant and time-consuming. Even determining the right source variant for cloning, or determining the target variant of a change propagation, is challenging, since it is often far from obvious what features are realized in a variant. Diffing complex software assets between variants is usually not meaningful.}
A scalable strategy is to integrate the cloned variants into a \emph{configurable and integrated platform}, by adopting platform-oriented engineering methods, such as software product line engineering (SPLE)\,\cite{clements.ea:01:software,czarnecki.ea:00:generative,linden.ea:2007:practices,apel2013software,berger2020state}. Individual variants are then derived by configuring the platform. This strategy is typically advocated for systems with many variants, such as software product lines (e.g., automotive/avionics control systems and industrial automation systems) or highly configurable systems (e.g., the Linux kernel). This strategy scales, but is often difficult to adopt and requires substantial up-front investments, since variability concepts (e.g., a feature model\,\cite{kang.ea:1990:foda,damir2019principles}, feature-to-asset traceability\,\cite{berger2010featuretocode,linsbauer2013recovering}, a configuration tool\,\cite{bashroush2017case}) need to be introduced and assets made reusable or configurable.
\longv{These concepts include, among others, features, code-level configuration (e.g., using a preprocessor), feature-to-asset traceability, a feature model (a tree-like representation of features and their dependencies\,\cite{kang.ea:1990:foda,damir2019principles}), and a configurator tool.}
In practice, organizations often start with \co and later face the need to migrate to a platform in a risky and costly process\,\cite{Stallinger:2011:MTE:1985484.1985490,Jepsen2007,berger2013survey,assunccao2017reengineering}, recovering meta-data that was never recorded during \co, such as features and their locations in software assets\,\cite{jacob2020apogamesmigration,damir2019principles}.

\longv{In summary, organizations need to decide for one of these two opposing strategies and often perform risky and costly re-engineering efforts. In this light, is it possible to exploit a spectrum between \co and platform-oriented engineering? Can we provide methods and tools that allow to seamlessly and incrementally transition between both opposing strategies? Can we continuously record meta-data and exploit it for the transition? 
}

\looseness=-1
Over the last decades, researchers focused on heuristic techniques to recover information from legacy codebases, including feature identification\,\cite{variclouds,zhou2018identifying,bennasr17}, feature location\,\cite{Rubin2013,dit2013feature,michelon2021hybrid}, variability mining\,\cite{kastner2011variability,kastner2013variability}, and clone-detection techniques\,\cite{roy2007survey,rattan2013software}. Unfortunately, such techniques are usually not accurate enough to be applicable in practice, and also require substantial effort to set them up and provide with manual input (e.g., specific program entry points for feature location techniques\,\cite{wang.ea:2013:howfeaturelocationdone}). \edit{As we will show, existing platform migration techniques either heavily rely on such heuristics or have only been formulated as abstract frameworks so far. Moreover, they tend to prescribe non-iterative, waterfall-like
migrations, making it risky and expensive.}

\looseness=-1
We take a different route and present a method to continuously record the relevant meta-data already during \co, and to incrementally transition towards a more scalable platform-oriented strategy, exploiting the meta-data recorded.
We design, formalize, and prototype a lightweight method called \textit{virtual platform}, generalizing clone-management and product-line migration frameworks\longv{, also combining and extending ideas from other researchers\,\cite{rubin.ea:2013:framework,Rubin2015,neves2015safe,fischer.ea:2014:ecco,Linsbauer2017,ji2015maintaining}}.
We exploit a spectrum between the two extremes of ad hoc \co and fully integrated platform, supporting both kinds of development.
As such, the virtual platform bridges \co and platform-oriented development (SPLE).
Based on the number of variants, organizations can decide to use only a subset of all the variability-implementation concepts that are typically required for an integrated platform. This allows organizations to be flexible and innovative by starting with \co and then incrementally adopting the variability-implementation concepts necessary to scale the development, \edit{as indicated by industrial practices for product-line adoption\,\cite{gruner2020incremental,berger2020state,kruger2020promote,zhang2013variability}.}
This realizes an incremental adoption of platforms with incremental benefits for incremental investment. Furthermore, it also allows to use \co even when a platform is already established, to support a more agile development with cloning and quickly prototyping new variants.
The framework is lightweight, since it avoids upfront investments and can be easily integrated with version-control systems or IDEs, where its operators can be mapped to existing activities, avoiding extra effort.
This way, our new (feature-oriented) operators are cheap to invoke during development, when the feature knowledge is still fresh in the developer's mind, allowing to record meta-data in a lightweight way.

\looseness=-1
The term ``virtual platform'' was introduced earlier in a short paper\,\cite{antkiewicz2014flexible} discussing an incremental migration of 
 clone-based variants into a platform.
It introduced governance levels reflecting a spectrum between the two extremes ad hoc \co and fully integrated platform.
Higher levels involve a super-set of the variability concepts of lower levels. Advancing a level---e.g., when the number of variants increases---supports an incremental adoption of variability concepts, avoiding the costly and risky ``big bang'' migration\,\cite{Stallinger:2011:MTE:1985484.1985490} often leading to re-engineering efforts over years\,\cite{Jepsen2007,Fogdal2016}.
This early, high-level description of a strategy to incrementally scale the management of variants paved the way for this paper.

\looseness=-1
One of our core contributions are conceptual structures and formalized operators for the virtual platform, which are related to ordinary code editing, but also record and exploit meta-data. 
While we prototypically implemented the virtual platform on top of an ordinary file system, our work gives rise to realize it upon a database (to enhance scalability), within an integrated development environment (IDE), or as a command-line tool. The meta-data could also easily be saved directly in the software-assets using lightweight embedded annotations (as our prototype does).
\longv{In fact, going back to the 1970s, researchers have built so-called variation-control systems\,\cite{Linsbauer2017}, which never made it into the practice of software engineering. These systems have been realized upon different back- and frontends (e.g., version-control systems\,\cite{munch.ea:1993:uniformversioning,DBLP:conf/scm/LieCDK89} or a text editor\,\cite{DBLP:journals/ibmrd/Kruskal84}), but before effective and scalable concepts from SPLE research for managing variability have been established.}

We evaluated our prototype on a reasonably sized  system (57.4k lines of text, 4 variants), 
where we simulated evolution activities that are typical of practical software systems.
Our prototype was able to fully simulate and manage all considered activities.
From a cost-benefit analysis, we conclude that the virtual platform offers significant cost savings during inevitable evolution and maintenance activities.

\looseness=-1
In summary, we contribute:
\begin{itemize}
	\item
	\textbf{a mechanization} of the so-far abstract idea of operators mediating between \co and an integrated platform, defined upon conceptual, language-independent structures,
	\item
	\textbf{a prototype of the virtual platform\,\cite{appendix:impl}} in Scala,
	
	\item
	\textbf{a comparative evaluation of the virtual platform} against five related frameworks, based on their ability to support common evolution scenarios,
	\item
	\textbf{a cost-and-benefit evaluation of the virtual platform}, based on a simulation study featuring the revision history of a real variant-rich open-source system, and 
	
	\item
	\textbf{an online appendix\,\cite{appendix:online}} with a technical report about our operators, additional examples, and evaluation data.
	
\end{itemize}

\noindent
\section{Motivation and Overview}
\noindent
\looseness=-1
We provide a core scenario of seamless variability management as a running example and an overview of the virtual platform. While rooted in a deliberately simple application domain, the example is inspired by  documented real product-line migration scenarios \cite{assunccao2017reengineering,laguna2013systematic}.
It includes tasks that are tedious and error-prone in practice (e.g., bugfix propagation along branches).

\label{sec:motivation}

\subsection{Motivating Running Example}
\label{sect:calculatorexample}
\noindent
\looseness=-1
We now discuss relevant problems of managing variants inspired by actual industrial practices, also presenting our solution in the virtual platform and how a developer would use the virtual platform. Specifically, developers interact with the virtual platform by invoking its provided operators, either via the command-line or an integration with an IDE or version-control system provided by a tool vendor (see \secref{sec:overview} for details). While the traditional, asset-oriented operators can run transparently in the background, only the feature-oriented operators require an extra user interaction for invoking the operators. The operator are described in detail in \secref{sec:operators}.

\looseness=-1
Consider the scenario of an organization developing and evolving variants of a calculator tool. 
Our organization starts creating a project of a simple calculator called \textit{BasicCalculator} (BC) that supports basic arithmetics: \textit{addition}, \textit{subtraction}, \textit{multiplication}, and \textit{division}. Soon, based on customer requests, the organization needs to create variants of \textit{BC}, which have substantial commonalities, but also differ in functional aspects.

\begin{figure}[t]
	\centering
	\includegraphics[width=\columnwidth]{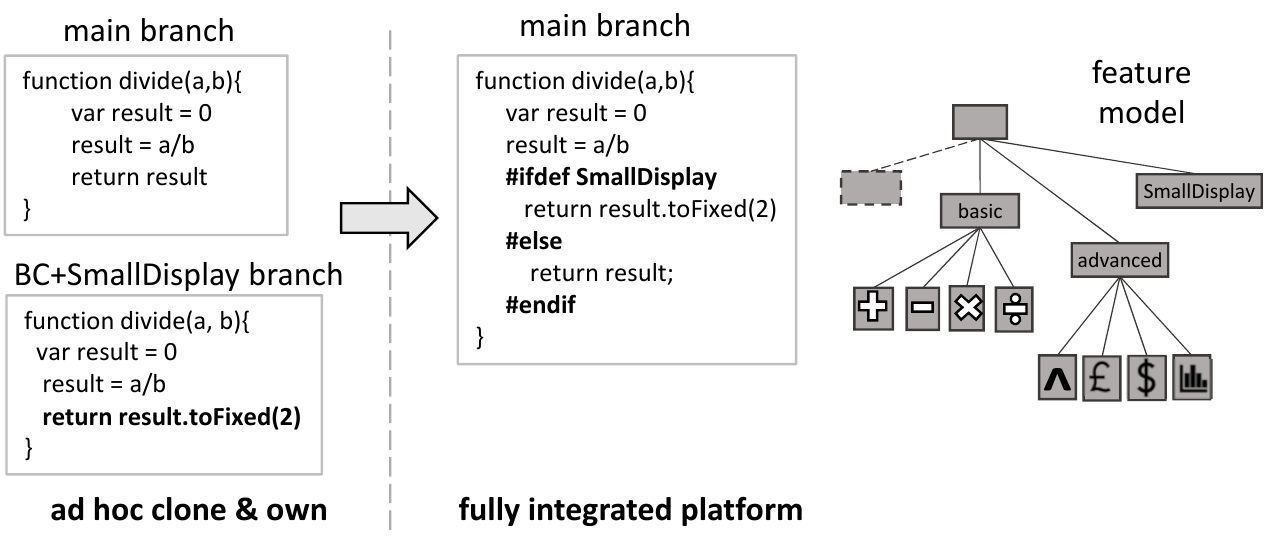}
	\vspace{-.8cm}
	\caption{Ad hoc \co vs. fully integrated platform illustrated for two variants: the \textit{BasicCalculator} and a variant with only a small display}
	
	\label{fig:extremes}
	\vspace{-.6cm}
\end{figure}

\looseness=-1
\Figref{fig:extremes} illustrates the two opposing strategies (cf. \secref{sec:introduction}) for realizing the variants.
Specifically, it shows two alternate realizations of a variant of \textit{BasicCalculator} with a small display, requiring the rounding of results (feature \textit{SmallDisplay}). To the left, the code is cloned and adapted (one line changed in the branch BC+SmallDisplay); to the right, a configuration option represents the change in a common codebase (integrated platform). The changes are usually more complex (e.g., features can be highly scattered\,\cite{passos.ea:2018:tse,passos.ea:2015:scattering}), as well as the representation of variability in the integrated platform. We also need more variability concepts, among others, features\,\cite{berger.ea:2015:feature,kruger2018towards,kruger2019my}, code-level configuration\,\cite{apel2013software}, feature-to-asset traceability\,\cite{berger2010featuretocode,linsbauer2013recovering,KruegerNF+17}, a feature model (a hierarchical structure with features and their dependencies)\,\cite{kang.ea:1990:foda,damir2019principles}, a configurable build system\,\cite{apel2013software}, and a configurator tool\,\cite{bashroush2017case,berger2013study,sincero.ea:osspl}. 
This example shows that, when it becomes necessary to migrate from \co to an integrated platform, important information needs to be recovered, specifically: that a feature \textit{SmallDisplay} was implemented and where its code is located. Recovering such information in systems with many features and sizable codebases is laborious, time-consuming, and inaccurate at best. Also, migration can be invasive, risky, and costly, especially hard to achieve under market pressure \,\cite{Hetrick2006,Jepsen2007,Fogdal2016,jacob2020apogamesmigration,Stallinger:2011:MTE:1985484.1985490,bilic.ea:2020:volvo}.

\begin{figure*}[t]
	\vspace{-.5cm}
	\centering
	\includegraphics[width=1.01\textwidth]{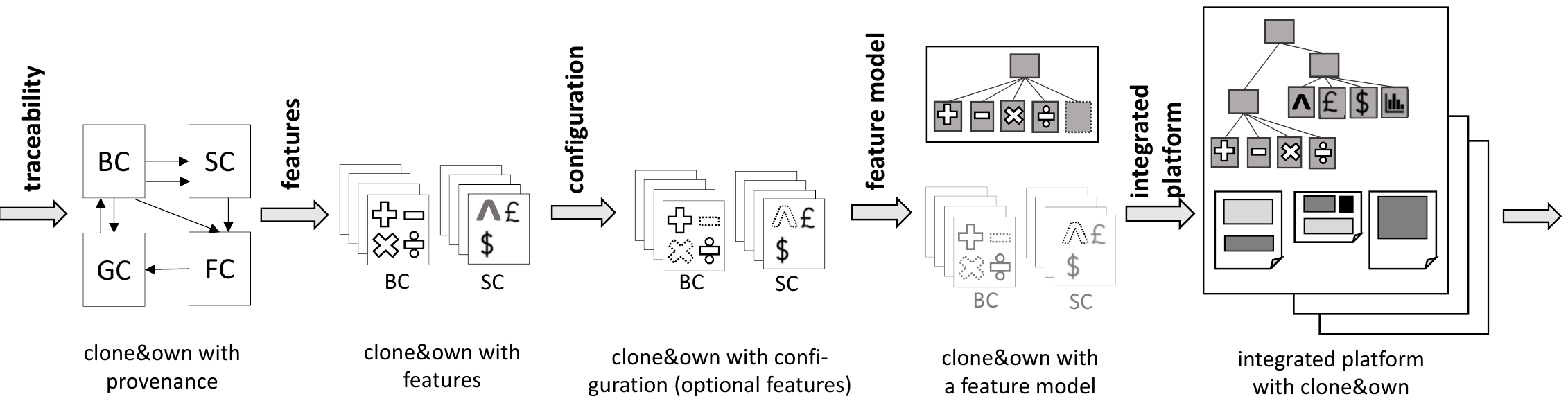}
	\vspace{-.8cm}
	\caption{Spectrum between the extremes ad hoc \co and a fully integrated platform (see \figref{fig:extremes} for both), illustrated with cloned variants: \textit{BasicCalculator} (BC), \textit{ScientificCalculator} (SC), \textit{GraphingCalculator} (GC), and \textit{FinancialCalculator} (FC). The virtual platform provides operators to transition along this spectrum (e.g., to incrementally adopt a platform).}
	
	\label{fig:spectrum}
	\vspace{-.4cm}
\end{figure*}

\looseness=-1
The virtual platform exploits a spectrum between the two extremes and supports an incremental transition as shown in \figref{fig:spectrum}. It adapts the governance levels from prior work\,\cite{antkiewicz2014flexible}, which also explains the benefits of each transition step in detail.

\looseness=-1
Let us further discuss the evolution of our calculator using \textbf{ad hoc \co}. After the \textit{BasicCalculator} and a variant of it for small displays (BC+SmallDisplay) is created, customers request a \textit{ScientificCalculator}, which should solve complex inputs, such as \textit{expressions}, \textit{factorials}, and \textit{logarithms}. Our organization decides to copy and adapt the codebase from \textit{BasicCalculator}, since there is no need for a \textit{ScientificCalculator} with small display support; otherwise we would already have four cloned variants.
As such, cloning provides a baseline minimizing the duplication of efforts. Soon after, the organization needs to create another variant called \textit{GraphingCalculator}, for which it selects the most similar variant, \textit{ScientificCalculator}, and clones and adapts it. It also notices that some functionality in \textit{BasicCalculator} had in the meantime received a bug fix, which the organization also applies to \textit{GraphingCalculator}, now realizing that also \textit{ScientificCalculator} needs to receive the bug fix.

\looseness=-1
\noindent
\textbf{Problem 1: Where are my clones?}
With many more variants developed using ad hoc \co, developers lose overview. If a change (e.g., a bug fix) is to be replicated, developers need to recover which project was cloned from which, in the worst case requiring a clone-detection technique. 
Also, the added effort in synchronizing cloned implementations is likely to surpass the initial benefit of reuse via cloning.

\looseness=-1
\noindent
\parhead{Solution 1: \Co with provenance} (\figref{fig:spectrum}, 1st level).
Our solution is to record traceability information about the cloned variants' provenance, which eases tracking and synchronizing clones. It also bypasses the inaccuracies associated with clone detection, making tasks such as change propagation more effective.
The virtual platform records clone traces among assets in the background, without requiring extra effort from the developer, but who can query it for obtaining the clones of an asset.

\looseness=-1
\edit{To this end, the developer invokes the \cloneasset operator provided by the virtual platform.
As a result, a trace between the original asset and its clone is stored in a trace database, which can be queried at any time by the developer to retrieve clones of an asset quickly and accurately. The developer can later propagate changes between the original asset and its clone (\propagateasset) or integrate changes between the assets (either manually or using a tool) by exploiting the continuously recorded meta-data. 
}

\looseness=-1
\noindent
\textbf{Problem 2: What is in my cloned variants?}
With more variants, despite provenance information, the problem arises that developers lose overview. To understand what is in the variants, we need a more abstract representation of assets. For cloning, this is also necessary to select an existing variant closest to the desired one in terms of the desired features. Furthermore, our organization finds the feature \textit{exponent} developed in \textit{ScientificCalculator} to be useful for other cloned variants. To clone it, the developer needs to know which implementation assets belong to the feature.

\looseness=-1
\noindent
\parhead{Solution 2: \Co with features} (\figref{fig:spectrum}, 2nd level). Add\-ing feature meta-data adds perspective and allows functional decomposition. It also allows representing assets in terms of features, to reuse and clone features across projects. Lastly, including feature-related information allows going past the efforts and inaccuracies of \textit{feature location} (recovering where a feature is implemented), making feature reuse and maintenance more effective. The virtual platform offers operators to add features conveniently (at the same time annotating assets).

\edit{The developer maps assets to features by using the operator \mapassettofeature. She can later query the virtual platform to find the location of the features using the operator \getmappedassets, and also to clone assets along with feature mappings (\cloneasset).}

\looseness=-1
\noindent
\textbf{Problem 3: How to reduce redundancy?}
Despite features, which help maintaining variants, substantial redundancy exists.

\looseness=-1
\noindent
\parhead{Solution 3: \Co with configuration} (\figref{fig:spectrum}, 3rd level).
To reduce it, 
our organization starts to incorporate configuration mechanisms. These allow to enable or disable features, such as \textit{SmallDisplay}, which control variation points. This reduces redundancy and maximizes reuse.
So, the organization maintains a list of features and uses a configurator tool.
The virtual platform supports this solution with a simple operator. 

\edit{Over time, the developer adds features by invoking the operator \addfeature. She can map the assets to features using \mapassettofeature and clone features using \clonefeature. She can also make features optional by invoking \makefeatureoptional. Variants can be configured by cloning the repository (\cloneasset) with assets mapped to only the selected features (\getmappedassets).}

\looseness=-1
\noindent
\textbf{Problem 4: How to keep an overview over the features?}
The more features and variation points the organization incorporates, the more it loses overview over the features and their relationships, including feature dependencies (accidentally ignoring those can lead to invalid variants). Maintaining such information would also help scoping variants.

\looseness=-1
\noindent
\parhead{Solution 4: \Co with a feature model} (\figref{fig:spectrum}, 4th level).
Our organization introduces a feature model, which captures features and their constraints, also as input to the configurator. Feature models are very intuitive and simple models, which provide deep insights without much additional tool support. They also foster communication among stakeholders and validate feature configurations. With this solution, consistency between features and clones is high, since developers can also exploit the clone traces and use the virtual platform for feature-based change propagation.

\looseness=-1
\edit{The developer adds a feature model to the repository with the operator \addfeaturemodeltoasset. She can change the feature model to add and remove features at any time. She can map assets to features from the feature model (\mapassettofeature), clone features across projects (\clonefeature), and propagate changes in features to their clones (\propagatefeature).}

\looseness=-1
\noindent
\textbf{Problem 5: How to keep consistency, improve quality, and further reduce redundancy?}
Our organization needs to further scale the development with an ever-increasing number of variants (due to rapidly changing market needs
), while it has problems maintaining consistency and propagating changes, despite some redundancy already being reduced with Solution 3. It is also likely that eventually, there will be some projects with a configuration mechanism and some without.

\looseness=-1
\noindent
\parhead{Solution 5: Integrated platform with \co} (\figref{fig:spectrum}, 5th level).
Our organization integrates the projects into a consolidated platform. Luckily it can exploit meta-data about clone traceability (\textit{provenance}) and features with their locations in assets. The virtual platform provides support for this kind of information, easing the integration of cloned variants into a platform. Of course, developers might have forgotten to record all that information, then it is natural to recover it. As long as some information is recorded, a benefit arises in terms of saved feature identification, feature location and clone-detection effort.

\noindent
\subsection{Virtual Platform Overview}
\label{sec:overview}
\looseness=-1
\noindent
Our goal is to combine the benefits of the two opposing strategies \co and integrated platform, exploiting a spectrum between both and allowing incremental transition as in our running example (\secref{sect:calculatorexample}). To this end, we designed a framework called virtual platform comprising conceptual structures upon which operators modifying the structures are executed by developers. The conceptual structures abstractly represent software assets at various levels of granularity---from whole repositories to blocks of code---and can be adapted to specific asset languages (explained shortly in \secref{sec:structures}).

In addition, they maintain information about variability, specifically feature information, feature-to-asset mappings, and clone traces.
\looseness=-1
The virtual platform extends other development tools, specifically, IDEs and version control systems.
On top of these, which are concerned with the management of \textit{assets}, the virtual platform provides dedicated functionality for managing \textit{features}.
Operators can be either traditional, meaning they are concerned with asset management, or feature-oriented, meaning they are devoted to features and their locations in assets. 
In contrast to traditional development workflows, the use of dedicated feature-oriented operators incurs a certain cost, but promises benefits to developers.
In \secref{sec:evaluation}, we study this trade-off.

\begin{figure}[b]
	\vspace{-.6cm}
	\centering
	\includegraphics[width=0.48\textwidth]{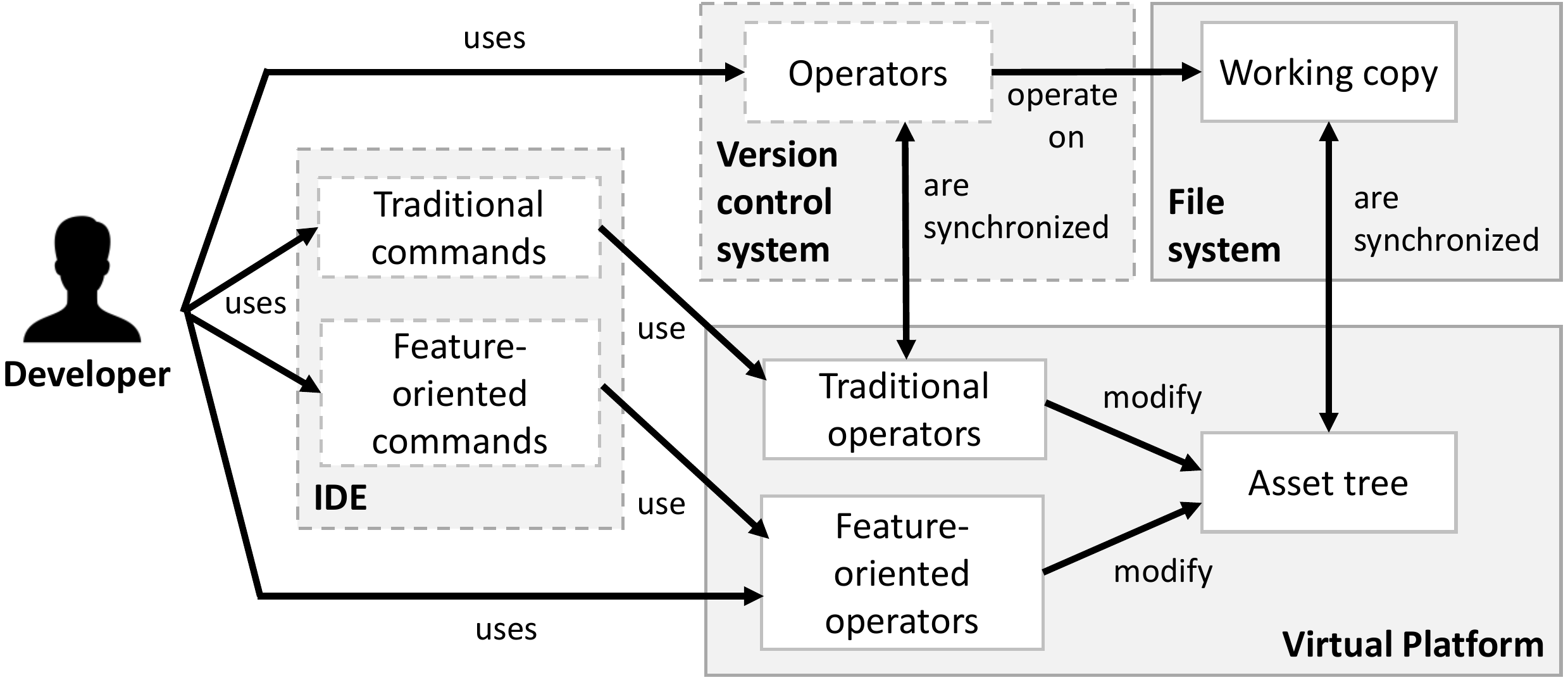}
	\vspace{-.3cm}
	\caption{Overview (dashed boxes represent optional parts)}
	\label{fig:overview}
\end{figure}

\looseness=-1
\Figref{fig:overview} illustrates interactions and internal workings of the virtual platform. Developers can interact with it directly or indirectly.
The former is enabled via extensions and hooks of existing tools.
Specifically, traditional IDE commands such as \textit{``Create File''} and version-control commands such as \textit{``Add File''} are linked to the traditional, asset-oriented operators of the virtual platform (e.g., \textit{``Create Asset''}) and do not impose additional effort for developers.
Feature-oriented operators can be implemented by new, feature-oriented IDE commands (e.g., \textit{``Create Feature''}).
Direct interaction is enabled via a command-line interface, where developers can call feature-oriented operations such as \textit{``Create Feature''} directly.

%To this end, we main a working copy on the developer's local file system, which can be coupled to a repository using the version control system.

\section{Methodology}
\label{sec:methodology}
 
\noindent
\looseness=-1
We followed a design-science-like strategy to iteratively define the conceptual structures, the operators, and to evaluate them using unit tests representing common scenarios. Specifically, for the structures and operators, we aimed at maximizing the support for different scenarios from the literature and our own professional experience. The main challenge was to define adequate structures that, while programming-language-independent, can be mapped to many of the different asset types of real-world software projects, as well as to design the operators to be able to operate on the structures.

\looseness=-1
\parhead{Initial Design.}
We started by analyzing clone-management and platform-migration frameworks proposed in the literature, from which we extracted development activities that should be supported by the virtual platform.
We also had a series of discussions among the authors, one from industry and four from academia. Two authors have over ten years of research experience in variability management and SPLE. We also created ad hoc examples in the discussion meetings. From these sources, we identified an initial set of data structures and operators, and implemented them in Scala.

Specifically, from the literature, we identified five relevant works on clone management and product-line migration using our expert knowledge. 
Rubin et al.'s product-line migration framework\,\cite{rubin.ea:2013:framework,Rubin2015} offers operators that support the narrative that a mechanization---i.e., an operator-based perspective---leads to more efficient implementation and support. Fischer et al.'s\,\cite{fischer.ea:2014:ecco} framework and tool ECCO relies on heuristics to identify commonalities and allows composing new product variants using reusable assets. Martinez et al.'s tool BUT4Reuse\,\cite{MartinezBut4Reuse} is an extraction-based technique for product-line migration, including support for feature-model synthesis. Pfofe et al.'s tool VariantSync\,\cite{variantsync:2016} supports clone-management by easing the synchronization of assets among cloned variants. Montalvillo et al.'s operators and branching models for clone management in version-control systems\,\cite{montalvillo2015tuning} allow isolated variant development with change propagation, but without using the notion of features, as opposed to the other frameworks.
For brevity, we will present the identified activities only at the end in \secref{sec:compeval}. Detailed descriptions are in our online appendix\,\cite{appendix:online}.

\looseness=-1
\parhead{Continuous Evaluation.}
Once every operator was implemented, we tested it with unit tests
based on scenarios from the literature and our own experiences.
We ensured that the operators assured the \wellformedness\ of the conceptual structures by prohibiting illegal actions, e.g., limiting asset addition to scopes that can host an asset of the given type.

\looseness=-1
\parhead{Final Qualitative and Quantitative Evaluation.}
We evaluated the virtual platform qualitatively by comparing it against the existing frameworks discussed above, from which we had extracted activities supported by techniques for supporting \co or the migration of cloned variants to an integrated platform. We evaluated the virtual platform quantitatively in a cost-benefit calculation based on simulating the development of a real open-source system developed using \co.

\section{Conceptual Structures}\label{sec:structures}
\begin{figure}[t]
	\vspace{-.3cm}
	
		\includegraphics[width=1.\columnwidth]{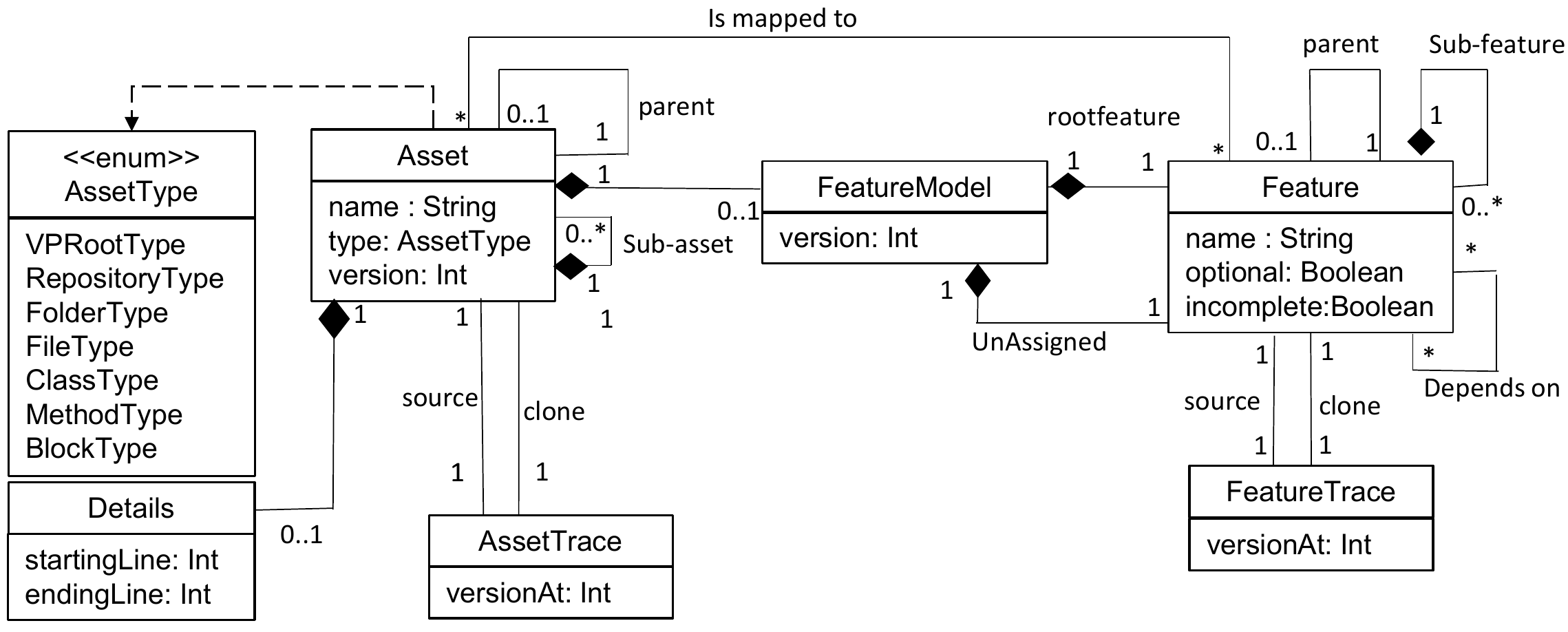}
	
	\vspace{-.4cm}
	\caption{Conceptual structures: asset tree, features, mappings, and clone traces}
		
	\label{fig:aststructure}  
	\vspace{-.6cm}
\end{figure}

\looseness=-1
\noindent
The virtual platform's conceptual structures form the basis for its operators, which we formulated as functions with side effects (in-place transformations) that modify the structures. \Figref{fig:aststructure} illustrates the main structures and their relationships. We define them abstractly, but also provide a concrete implementation for handling assets within a file system and special support for textual files that follow a hierarchical structure (e.g., with nested classes, methods or code blocks; cf. \secref{sec:implementation}).

\looseness=-1
\parhead{Asset Tree ($AT$)}
is our main conceptual structure and abstractly represents a hierarchy of assets, such as the folder hierarchy, but also the hierarchy within source files. \edit{In \figref{fig:aststructure}, the AT is represented implicitly in the form of assets with their sub-asset relationships.}
The idea of $AT$ is inspired by feature structure trees (FSTs,\,\cite{apel.ea:2009:featurehouse}), which represent source files. In our case, we define the $AT$ as a hierarchical, non-cyclic tree structure of nodes.
It has a synthetic root node (\rootnode) and then represents a hierarchy that can start with repositories as the top-level nodes, followed by folders and files, and can then go into the nesting structure of elements of hierarchical files. 

Every node represents an
\asset related to the project, such as a folder, a file (e.g., image, source file, model or requirements document), or text. Every \asset has a \name, a type (\assettype), and a \version (a simple means to identify changes). An asset can have any number of \subassets. It also owns a \parent pointer $p$, which should define a tree, with a virtual root node (\asset of type \vproot) denoted as \rootnode.
The \assettype is used to capture the role of the \asset in the project, and can be one of the following: \vproot, \repository, \folder, \file, \class, \method, and \block.
The type \vproot is only used once in the $AT$, to specify the synthetic root node.
The main purpose of this root node is to carry a global version (we explain versioning shortly).

\looseness=-1
Traditional SPLE architectures have a feature model per project, which can be difficult to maintain and evolve in large systems (e.g., Linux kernel\,\cite{she2010variability}).
We provide a more flexible structure by including an optional feature model as part of every \asset (see composition of \featuremodel in \asset in \figref{fig:aststructure}).

\looseness=-1
\parhead{Well-Formedness Criteria}
We define a partial order of valid containment over the types of assets in a check function $\containable: \asset\times\asset\rightarrow\bool$ that validates the containment based on the asset types. For instance, \vproot can only be at the root, and a \method can be contained in a \file, but not the other way around. Operators are implemented with consideration of \wellformedness\ criteria, to ensure that the tree structure of $AT$ is retained.

\looseness=-1
\parhead{Features and Feature Models}
A \featurecmd\ has a \name and two \texttt{Boolean} parameters: \optional and \incomplete. The field \optional specifies whether the \featurecmd\ is mandatory or optional; \incomplete captures information about the completeness of the \textsl{feature's} implementation.
If the \featurecmd\ was cloned from another \featuremodel\ scope, it is \texttt{true} if the new scope containing the \featurecmd\ also contains all the assets to which the \featurecmd\ is mapped; otherwise it is always \texttt{false}.
Every \featurecmd\ has an optional \parent, and any number of \subfeatures. Features can have dependencies to each other.

A \featuremodel\ has a \textit{root} \featurecmd\
and a mandatory \featurecmd called \Unassigned, which contains all features that are added to the model as a result of asset cloning. That is, 
if any \featurecmd\ mapped to the \asset is not present in the target \featuremodel\ already, it is mounted under \Unassigned (and requires developer intervention to move it to the desired location in the model).

\looseness=-1
\parhead{Asset-To-Feature Mappings},
in practice, can have two semantics. They can be simple mapping 
relationships, indicating that \asset realizes a \featurecmd \cite{ji2015maintaining}. They can also indicate variability \cite{struber2020variability}, where the \asset is included in a concrete variant if the \featurecmd\ is selected (interestingly, if an \asset is optional based on a \featurecmd, then the \asset also realizes it, but not necessarily all assets realizing a \featurecmd\ are optional). The SPLE community usually focused on the variability relationship, and the feature-location community on traceability. For the virtual platform, we unified the mechanism with which assets are mapped to features. Specifically, an \asset has a presence condition (PC)---a propositional formula over features. A PC allows conveniently mapping assets of different granularity levels (\assettype) to entire \featurecmd\ expressions.
Whether this relationship to the \featurecmd\ represents variability or traceability is solely determined by the \textsl{feature's} \optional parameter.

\looseness=-1
\parhead{Versioning of Assets.}
Assets (and features) have a \version---an integer used to recognize changes in the $AT$ (and $FM$), especially among cloned assets.
The \version of the \vproot node has a special role, which we call ``\globalversion'' and which carries the most up-to-date \version, to recognize any change in the whole $AT$.
For simplicity, we assume that any \asset outside the tree has a \version of 0.
After addition, it takes the \version of the global root (initialized with 1 and incremented after any update in the $AT$). Versions are incremented after every modification and addition or removal of \subassets. 
This simple versioning strategy is a sweet spot between two other alternatives:
First, after every change in an \asset, increment the \version of the \asset and continue updating the ancestors up to the root. This would make the tracking of the changes easy, but change propagation expensive and redundant. 
Second, keep two separate numbers, one global version, and one local version for every asset. This solution would ease change propagation, but yield a hard-to-understand versioning model.

\looseness=-1
\parhead{Clone Traceability.}
To maintain trace links between source assets and their clones, we define an \tracedb---essentially a list of \assettraces (\figref{fig:aststructure}).
An \textsl{AssetTrace} is a triplet of the source \asset, its clone, and a  \version at which the source \asset was cloned. Similarly, \featurecmd\ traces are used to keep track of the \featurecmd clones, and they are stored in a \ftracedb. A \textsl{FeatureTrace} is also a triplet pointing to the source \featurecmd, its clone, and  \version at the time of cloning. These traces are a core component of our contribution, and have special relevance in cloning and change propagation for both assets and features. For brevity, we refer to both \tracedb and \ftracedb as \textsl{TraceDatabase} in the remainder of the paper. %The variable \textit{VersionAt}\tb{necessary here? don't really understand the sentence} is very pertinent during change propagation, where it is compared against the current \version of the source to determine if a change propagation is due.

\noindent
\section{Virtual Platform Operators}\label{sec:operators}
\noindent
\looseness=-1
We now present the traditional, asset-oriented and the feature-oriented operators. Their underlying algorithms and further illustrations (supplementary to the illustrations used here) are provided in our online appendix\,\cite{appendix:online}. The appendix also presents a number of additional \textit{convenience operators}---utility methods that efficiently traverse the trees ($AT$ and \featuremodel) to return data that needs to be frequently accessed (such as assets mapped to a \featurecmd\ and clones of an \asset\ etc).

\newcommand{\descr}{\csname @beginparpenalty\endcsname100000 
\noindent{}\textit{Description:}}
\newcommand{\precond}{\noindent{}\textit{Precondition:}}
\newcommand{\preconds}{\noindent{}\textit{Preconditions:}}
\newcommand{\postcond}{\noindent{}\textit{Postcondition:}}
\newcommand{\postconds}{\noindent{}\textit{Postconditions:}}
\newcommand{\examp}{\noindent{}\textit{Example:}}
\newcommand{\versioning}{\noindent{}\textit{Versioning:}}
\longv{
\subsection{Versioning Strategy}\label{sec:versioningstrategy}
We designed a simple
version strategy, aimed at finding a sweet spot between two other possible solutions:

First, after every change in an \asset\ (e.g., addition of sub-asset,
deletion or modification), increment the \version of the \asset\ and continue updating the ancestors up to the root. This makes the tracking of the changes easy, but change propagation is expensive and redundant. 
Second, keep two separate numbers, one global version, and one local version for every asset. This solution makes change propagation easier, but leads to a more hard-to-understand versioning model.
Our final strategy
 
is that \vproot has a \globalversion. After any update in the $AT$, this number is incremented and all affected assets (added, parented, changed, cloned) get it. For change propagation, instead of a full tree-diff, the \tracedb is traversed to determine if there were any changes after the clone. Lastly, note that \vproot is a virtual \asset\ primarily used for versioning, and does not essentially translate to an actual artifact in the code base. For feature models, there is a \globalversion per \featuremodel\, which is incremented after any change in the \featuremodel\. For detecting changes between features (source and target features for change propagation), there are a few \textit{convenience operators} which compare their properties and structure to determine whether they have been modified or not.
}
\subsection{Traditional/Asset-Oriented Operators}\label{sec:operators:assetoriented}
\begin{figure}[b]
	\centering
	\vspace{+.3cm}
	\includegraphics[width=.9\columnwidth]{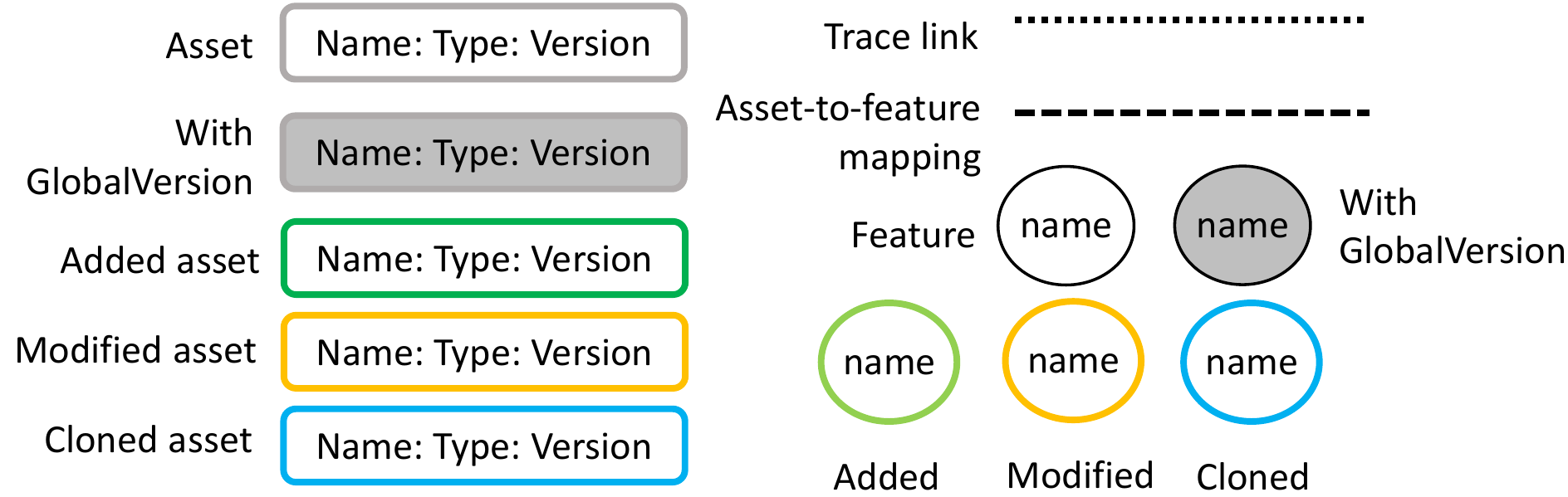}
	\vspace{-.3cm}
	\caption{Notations used in operator illustrations}
	\label{fig:legend}
\end{figure}

\noindent
\looseness=-1
We represent conventional activities performed by developers using asset-oriented operators. These operators allow to keep the $AT$ in sync with the working directory. Also, the assets act as \textit{mappable} components to the features, and allow cloning and change propagation.
In what follows, we introduce the asset-oriented operators with their parameter types, a brief description, and sample scenarios, inspired from our calculator running example (cf. \secref{sect:calculatorexample}). 
The notation used for visualizing various scenarios is shown in \figref{fig:legend}.

\noindent{}$\textbf{\addasset}:\asset\times\asset\rightarrow\bool$

\descr\ When a source \asset\ ($S$) is added in any target \asset\ ($T$) to a repository (e.g., a file to a folder), \addasset\ creates an \asset\ for $S$ and adds it to the preexisting \asset\ $T$ in the $AT$. Additionally, it increments the \globalversion, and assigns it to $S$ and $T$. This implies that the most recently changed assets are $S$ and $T$. Also, it adds any \featurecmd\ mapped to $S$ in T's \featuremodel\ (typically repository \featuremodel).

\begin{comment}
	\precond\ S satisfies the \wellformedness\,criteria with respect to T; \edit{source is \containable\,in target}.
	
	\postconds\ (1) S's \parent is set to T\,
	(2) \globalversion is incremented and assigned to S and T\,
	(3) All mapped features of S are added in the feature model of T.
\end{comment}

\begin{figure}[t]
	
	\centering
	\includegraphics[width=.5\columnwidth]{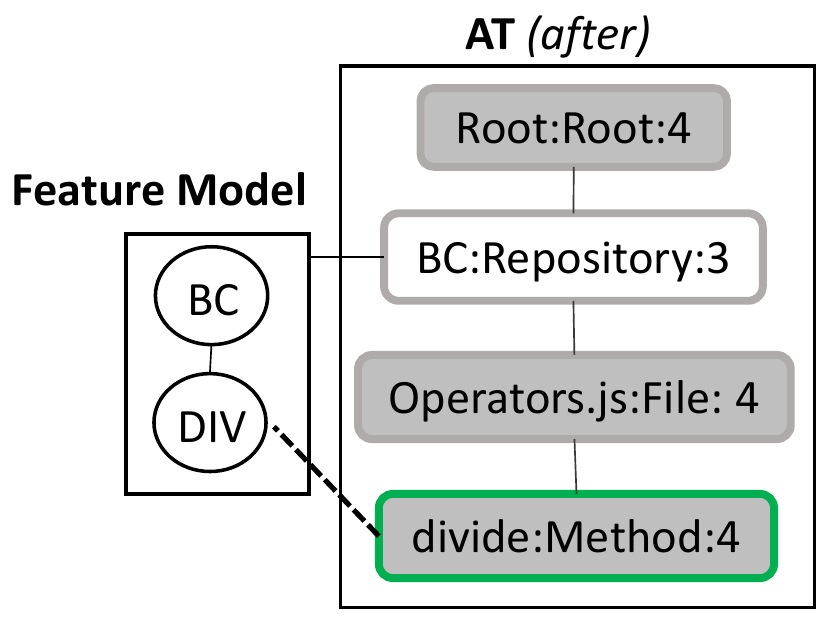}
	\vspace{-.4cm}
	\caption{Illustration of \addasset{(\textit{divide},\textit{Operators.js})}}
	\label{fig:addasset}
	\end{figure}

\examp\ Consider the \textit{BasicCalculator (BC)} example. The developer adds the implementation for the \textit{divide} method in the file \textit{Operators.js}, with an annotation for the \featurecmd\ \textit{DIV}. Consequently, the virtual platform creates and adds the \asset\ \textit{divide}\,(S) of \method\ to the \asset\ \textit{Operators.js}\,(T) of \file, and \textit{DIV} to the \featuremodel\ of $T$. The \globalversion (previously 3) is incremented and assigned to \textit{divide}\,and \textit{Operators.js}. \Figref{fig:addasset} illustrates the scenario.

\newlength{\textfloatsepsave} \setlength{\textfloatsepsave}{\textfloatsep} 
\setlength{\textfloatsep}{0pt}

\algo{
\begin{algorithm}[h]
\caption{Operator \addasset( S, T )}
\label{alg:addasset}
\fontsize{7}{7.5}\selectfont
\begin{algorithmic}[1]
\Require \containable(S, T)
\State S.parent = T
\State T.children += S
\State \updateversion(\rootnode, S, T)
\For{ $\textit{mf} \leftarrow S.\mappedfeatures$}
\State \add( mf, T.featuremodel )
\EndFor
\end{algorithmic}
\end{algorithm}
\setlength{\textfloatsep}{\textfloatsepsave}

\smallskip}
\looseness=-1
\noindent{}$\textbf{\changeasset}:\asset\rightarrow\bool$

\descr\ Upon a change in an \asset\ $S$ in the repository, \texttt{ChangeAsset}\xspace  increments the \globalversion of the $AT$ and assigns it to $S$. Versionable changes include renaming, addition, mapping to a \featurecmd\ and modification or removal of lines.

\noindent{}$\textbf{\removeasset}:\asset\rightarrow\bool$  

\descr\ If an \asset\ is deleted from a parent asset $T$, \removeasset removes its corresponding \asset\ $S$ in the $AT$, along with all its sub-assets. It increments the \globalversion and assigns it to $T$. Additionally, any \featurecmd\ mapped to $S$ is also removed from the \featuremodel\ of $S$ if $S$ the \textit{only} \asset\ mapped to it. This enforces that if all assets mapped to a \featurecmd\ are deleted, the \featurecmd\ is also deleted.

\begin{comment}
	\precond\ T is the parent of S. 
	
	\postconds\ (1) T does not have S as a sub-asset anymore
	(2) The \parent of S is nullified
	(3) The \globalversion is incremented and assigned T. 
\end{comment}

\noindent{}$\textbf{\moveasset}:\asset\times\asset\rightarrow\bool$

\descr\ If an \asset\ is moved from one location to another, \moveasset{} clones the corresponding \asset\ $S$ to the new target \asset\ $T$ (using \cloneasset), and removes it from the sub-assets of its previous parent (using \removeasset). 

\begin{comment}
\preconds\ S and T belong to the same AT.

\looseness=-1
\postconds\ S's previous \parent does not have S as a sub-asset.

\versioning\ The operator is a combination of \removeasset{}, and \addasset{}. The versioning is also the same; the former's versioning is adopted for S's previous parent, and the latter for T.
\end{comment}
\noindent
Thus far, the operators we presented serve two purposes: keeping the $AT$ synchronized with the project, and keeping track of changes through versioning. Following, the operators serve two additional purposes: storing feature-oriented data, and recording traceability among clones. The exploitation of these meta-data are the essence of our framework.

\noindent{}$\textbf{\mapassettofeature}:\asset\times\featurecmd\rightarrow\bool$

\descr\ Upon addition of a \featurecmd\ mapping by a developer, \mapassettofeature{} checks if the \featurecmd\ exists in the \featuremodel\ of the \asset. If not, it creates a \featurecmd\ $F$ (with the name used by the developer), maps it to $S$ (corresponding \asset\ in the $AT$), and adds $F$ to the \Unassigned\ \featurecmd\ in the \featuremodel\ of $S$. If $F$ already exists, it simply maps $F$ to $S$. For mapping, it adds $F$ to the \presencecondition\,of $S$ with a logical disjunction. To track this change, the \globalversion is incremented and assigned to $S$.

\begin{comment}
	\preconds\ (1) The asset belongs to the AT (2) The feature exists in the closest feature model of S (feature model belonging to the closest ancestor of S having a feature model).
	
	\postconds\ (1) The feature is added to the \presencecondition of the asset (2) \globalversion is incremented (3)  The
	mapped asset has acquired \globalversion.
\end{comment}

	\begin{figure}[b]
	\centering
	\includegraphics[width=.7\columnwidth]{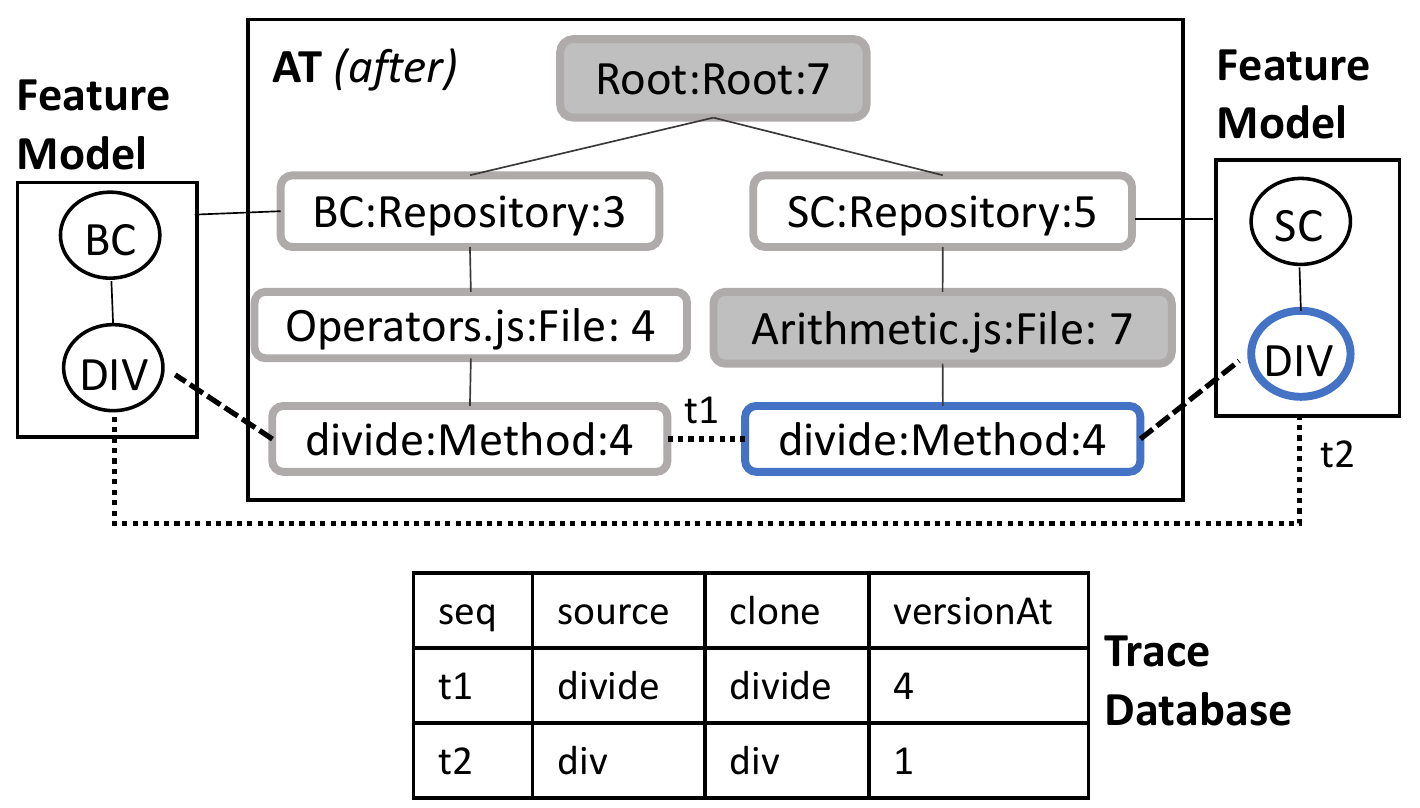}
	\vspace{-.2cm}
	\caption{Illustration of \cloneasset{(\textit{divide}, \textit{Arithmetic.js})}}
	
	\vspace{-.3cm}
	\label{fig:cloneasset}
\end{figure}

\examp\ Assume that the developer adds a method \textit{multiply} to \textit{BC}, with a \featurecmd\ annotation for the \featurecmd\ \textit{MULT}. \mapassettofeature creates this mapping in the $AT$. The \presencecondition of the method becomes ``\textit{MULT} $|$ \texttt{true}''.

\algo{\setlength{\textfloatsep}{0pt}
	\begin{algorithm}[h]
		\caption{Operator \mapassettofeature ( S, F )}
		\label{alg: mapassettofeature }
		\fontsize{7}{7.5}\selectfont
		\begin{algorithmic}[1]
			
			\State A = getAncestorWithFM(S)
			\If{ FMContainsFeature(A.featureModel,F) == false}
			\State AddUnassignedFeature(A.featureModel,S)
			\State updateversion(A.featureModel)
			\EndIf
			\State S.PC = F | S.PC
			\State updateversion(root,S)
		\end{algorithmic}
\end{algorithm}}

\noindent{}$\textbf{\cloneasset}:\asset\times\asset\rightarrow\bool$

\descr\ \cloneasset imitates the actual \co\ strategy; when an \asset\ is cloned to another location by a developer, \cloneasset creates a \textit{deep} clone of the source \asset\ and adds it to the target \asset\ in the $AT$, provided it is \containable. Additionally, if the cloned \asset\ (or its sub-assets) is mapped to any features, they are also cloned, added to the target \featuremodel, and mapped to the \asset\ clone. The clone retains the \version of the original asset, however, since the target \asset\ is modified (addition of sub-asset), the \globalversion is incremented and assigned to the target. For storing trace links, it creates traces for both \asset\ and \featurecmd\ clones and adds them to the \textsl{TraceDatabase}.

\begin{comment}
	\precond\  (1.) S is \containable in T.
	
	\looseness=-1
	\postconds\ (1.) A clone of S is added to the T; 
	(2.) A trace for every asset from the deep clone to its original is recorded;
	(3.) The mapped features of source are cloned in target and mapped to cloned assets; 
	(4.) A trace for every cloned feature to its original is recorded;
	(5.) All asset clones have the same version number as their original.
	(6.) \globalversion is updated. 
	(7.) The target's \version number is set to the new \globalversion.
\end{comment}

\begin{comment}
\begin{figure*}[ht]
\begin{minipage}[b]{1.0\linewidth}
\centering
\vspace{-.3cm}
\includegraphics[width=.25\linewidth]{fig/addasset.pdf}
\label{fig:1}
\includegraphics[width=.40\linewidth]{fig/cloneasset.pdf}
\label{fig:2}
\includegraphics[width=.30\linewidth]{fig/propagateasset.pdf}
\label{fig:3}
\end{minipage}
\caption{Testing}
\end{figure*}
\end{comment}

\examp\ Starting from \figref{fig:addasset}, the developer copies the method \textit{divide} in \textit{Arithmetic.js}; a file in another project, \textit{ScientificCalculator (SC)}. \cloneasset clones \textit{divide} to \textit{Arithmetic.js}, an \asset\ of \file in \textit{SC}, as well as the mapped \featurecmd\ \textit{DIV} in the \featuremodel\ of \textit{SC}. Traces for both \textit{divide} and \textit{DIV} are added to the \textsl{TraceDatabase}. \Figref{fig:cloneasset} illustrates the scenario.

\algo{
\setlength{\textfloatsep}{0pt}
\begin{algorithm}[h]
	\caption{Operator \cloneasset(S, T )}
	\label{alg:cloneasset}
	\fontsize{7}{7.5}\selectfont
	\begin{algorithmic}[1]
		
	\State C = S.Clone
	\State C.PC = True()
	\State \addasset(C,T)
	\State FM =  \getAncestorFeatureModel(T)
	\State mappedfeatures = S.\mappedfeatures
	\For {$\textit{mf} \leftarrow  S.\mappedfeatures$}
	\If  {!\featureexists(mf.name,FM) \& !\featurecloned(mf,FM)}
	\State FC = mf.Clone
	\State \addunassignedfeature(FC,FM)
	\State \mapassettofeature(C,FC)
	\Else 
	\State FC = \getfeature(mf.name,FM)
	\State \mapassettofeature(C,FC)
	\EndIf
	\EndFor
	\State \addtrace(S,C,S.version)
	\For {$\textit{child} \leftarrow S.subassets$}
	\State  \cloneasset(child,C)
	\EndFor
\end{algorithmic}
\end{algorithm}
\setlength{\textfloatsep}{\textfloatsepsave}
}

\noindent{}$\textbf{\propagateasset}:\asset\times\asset\rightarrow\bool$

\descr\ \propagateasset{} takes two assets, checks if one is a clone of the other, and propagates changes in source, after cloning, to its clone. To determine if source was changed, it compares the \version\ of source to its \version\ when it was cloned (\textsl{versionAt} from the \textsl{TraceDatabase}). If it is ahead of the \version\ it was cloned at, the changes are propagated to the clone. Changes performed in the clone are retained. Propagation, like cloning, includes added and modified sub-assets, added mappings, and renaming.
After propagation, a trace with source and clone is added to the \textsl{TraceDatabase}, the \textsl{versionAt} of which is the \version\ of the source. 
The \globalversion is incremented and assigned to the clone.

\begin{comment}
\preconds\ (1.) The target asset is a clone of source (identified via a clone trace). (2.)  The source's \version is ahead of the \version when it was cloned.

\postconds\ (1.) All modifications in source after cloning are propagated to target. (2.) The \globalversion is incremented and assigned to the target (3.) The trace database is updated.
\end{comment}

\begin{figure}[t]
	\centering
	\includegraphics[width=.5\columnwidth]{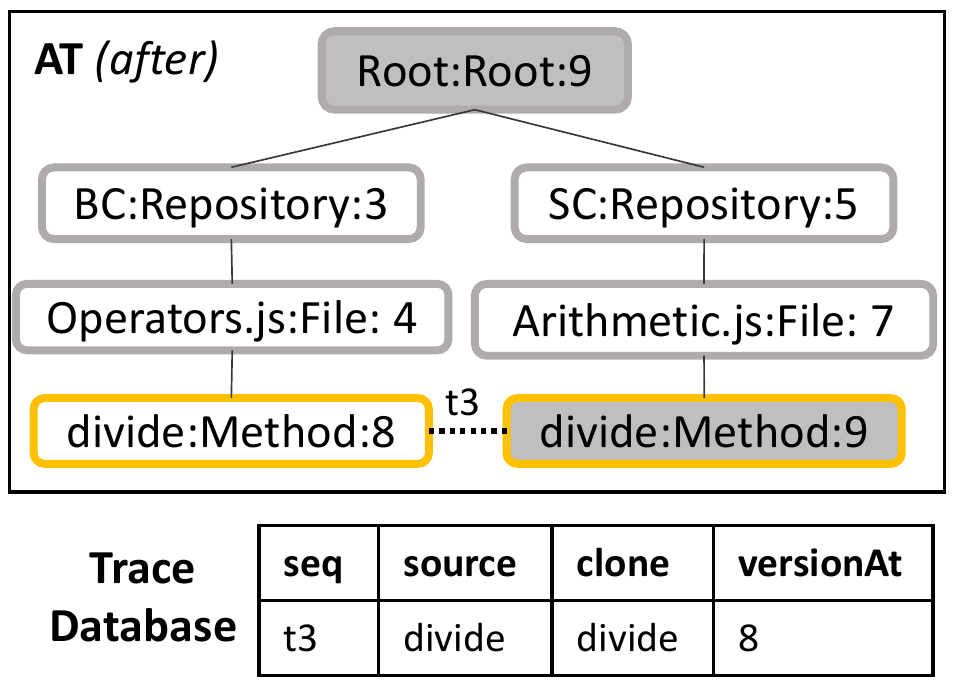}
	\vspace{-.3cm}
	\caption{Illustration of \propagateasset{(\textit{divide}, \textit{divide})}}
	
	\label{fig:propagateasset}
	
\end{figure}

\examp\ Assume that the \textit{divide} method during cloning did not include the check for division by zero. After adding the check (\changeasset), the \textit{divide} method in source (\textit{Operators.js}) is ahead (\textsl{version}=8) of the \textit{divide} method in target (\textit{Arithmetic.js}), with \textsl{version}=4. By invoking \propagateasset, the changes are propagated automatically. \Figref{fig:propagateasset} demonstrates the scenario; for simplicity, \featurecmd\ mappings are omitted.
\vspace{-0.04cm}

\algo{
\setlength{\textfloatsep}{0pt}
\begin{algorithm}[h]
	\caption{Operator \propagateasset( S, T )}
	\label{alg:propagateasset}
	\fontsize{7}{7.5}\selectfont
	\begin{algorithmic}[1]
		
	\State Trace = \getlatesttrace(S,T)
	\If{Trace.exists \& T.versionAt > S.version}
	\State \makeconsistent(S,C)
	\State FM = \getAncestorFeatureModel(T)
	\For {$\textit{mf} \leftarrow S.\mappedfeatures$}
	\If {!\featurecloned(mf,FM)}
	\State mfc = mf.Clone
	\State \addunassignedfeature(mfc,FM)
	\State	\mapassettofeature(T,mfc)
	\EndIf
	\EndFor
	\For{$\textit{child} \leftarrow S.subassets$}
	\If {\assetcloned(child,T)}
	\State \propagateasset(child,\getclone(child,T))
	\Else
	\State \cloneasset(child,T)
	\EndIf
	\EndFor
	\EndIf
\end{algorithmic}
\end{algorithm}
\setlength{\textfloatsep}{\textfloatsepsave}
}
\subsection{Feature-Oriented Operators}\label{sec:operators:featureoriented}
\looseness=-1
\noindent
The feature-oriented operators incorporate feature-related information to the $AT$ and enable feature reuse and maintenance.

\noindent{}$\textbf{\addfeature}:\featurecmd\times\featurecmd\rightarrow\bool$

\looseness=-1
\descr\ When a developer adds a \featurecmd\ (e.g., in a text file or a database), or an \asset\ mapping to a \featurecmd\ which does not exist in the \featuremodel, \addfeature{} creates a new \featurecmd\ and adds it to the \featuremodel. It also adds any assets mapped to the \featurecmd\ using \addasset{}. Similar to versioning of \addasset in $AT$, 
\addfeature\ increments the \globalversion (\version of \textit{root} \featurecmd) and assigns it to the added \featurecmd.

\begin{comment}
	\preconds\ (1.) The target repository's AT has a feature model. (2.)  The target feature exists in the feature model.
	
	\postconds\ (1.) The feature has been added as a top-level feature to the target feature model, or as a sub-feature to the target feature. (2.) Each asset mapped to the feature is contained in the target repository's AT and is mapped to the feature clone. (3.) The \globalversion of the feature model is incremented.
\end{comment}

\begin{figure}[t]
	\centering
	\includegraphics[width=.6\columnwidth]{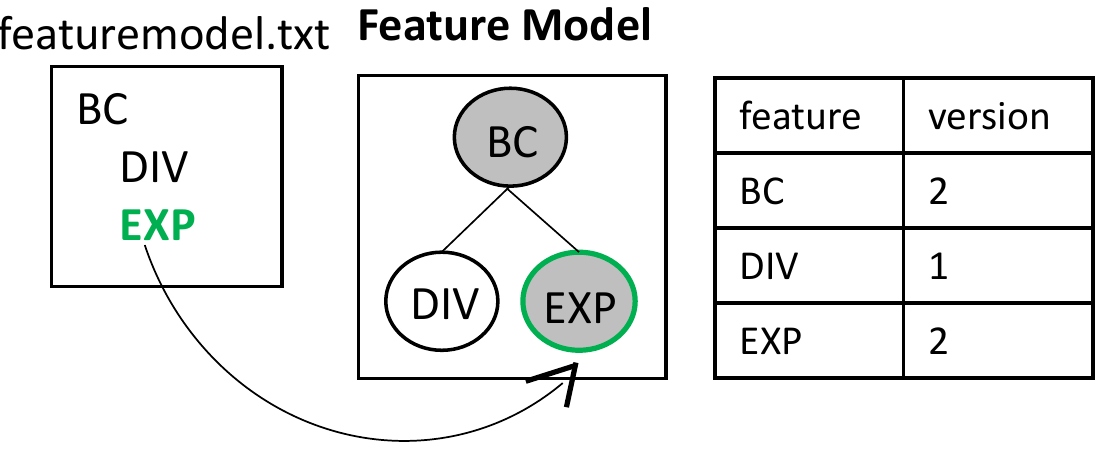}
	\vspace{-.3cm}
	\caption{Illustration of \addfeature{(\textit{EXP}, \textit{BC})}}
	
	\label{fig:addfeature}
	
\end{figure}

\examp\ Assume that the \featuremodel\ for \textit{BC} is a textual file, where features are written as individual lines, and indentation is used to represent hierarchy (Clafer syntax\,\cite{bkak2016clafer}). The developer adds a line ``EXP'' (exponent), below the line ``BC'' (\textit{root} feature, \textit{BC)}. \addfeature{} creates a corresponding \featurecmd\ \textit{EXP}, and adds it to the \featurecmd\ \textit{BC}. The \version\ of \textit{root} \featurecmd\ is incremented (previously 1 after adding \featurecmd\ \textit{DIV}) and assigned to \featurecmd\ \textit{EXP}. \Figref{fig:addfeature} demonstrates the scenario, with the resulting versions in a table on the right.

\noindent{}$\textbf{\addfeaturemodeltoasset:}\,\asset\times\featuremodel\rightarrow\bool$

\descr\ Developers can add a \featuremodel\ to an \asset\ in different ways, e.g., as a file or a database. The virtual platform, upon recognizing that a \featuremodel\ is added to an asset in the repository, invokes \addfeaturemodeltoasset. The operator then locates the \asset\ in the $AT$, creates a \featuremodel\ $FM$, and sets the asset's parameter \featuremodel\ to $FM$. The \globalversion of the $AT$ is incremented and assigned to the \asset\ which contains $FM$.

\begin{comment}
	\precond\ The asset belongs to the AT.
	
	\looseness=-1
	\postcond\ (1.) The asset's feature model is FM. (2.) \globalversion is incremented. (3.) The asset's \version gets the new \globalversion.
\end{comment}

\examp\ Consider that the \featuremodel\ of \textit{BC} is a separate text file, which resides in the root folder of \textit{BC}. As a result of \addfeaturemodeltoasset{}, the \featuremodel\ ($FM$) will be loaded from the file and assigned to \textit{BC}. All sub-assets of \textit{BC} can now be mapped to features from $FM$.

\noindent{}$\textbf{\removefeature}:\featurecmd\rightarrow\bool$

\descr\ When a \featurecmd\ is removed by a developer from a repository, \removefeature locates the \featurecmd\ in the \featuremodel, un-maps it from all assets it maps to, and removes the \featurecmd\ along with all its sub-features. Additionally, any asset mapped to \textit{only} the removed feature is also removed by the operator. The operator increments the \globalversion of the $FM$ and assigns it to the parent \featurecmd\ (before removal).

\begin{comment}
\precond\ The feature exists in the feature model. 

\looseness=-1
\postconds\ (1.) The feature is removed from the \subfeatures of its former \textit{parent} feature. 
(2.) The feature's \textit{parent} is null.
(3.) No asset mappings are maintained for the feature.
(4.) Any assets previously mapped only to the feature are removed from the AT.
(5.) The \globalversion is incremented.
(6.) All assets previously mapped to the feature and still in the AT acquire the \globalversion.
\end{comment}

\algo{
\setlength{\textfloatsep}{0pt}
\begin{algorithm}[h]
	\caption{Operator \removefeature(FP)}
	\label{alg:removefeature}
	\fontsize{7}{7.5}\selectfont
	\begin{algorithmic}[1]
		
	\State F = getFeature(SP)
	\State FM = \getAncestorFeatureModel(SP)
	\If{\featureexists(F,FM)}
	\For{$\textit{A} \leftarrow F.\mappedassets$}
	\If{A.\mappedfeatures.length == 1}
	\State \removeasset(A,\rootnode)
	\Else
	\State \unmapasset(A,SF)
	\EndIf
	\State \deletefeature(F,FM)	
	\State \updateversion(FM)
	\EndFor
	\EndIf
	\end{algorithmic}
	\end{algorithm}
	\setlength{\textfloatsep}{\textfloatsepsave}
}

\noindent{}$\textbf{\movefeature}:\featurecmd\times\featurecmd\rightarrow\bool$ 

\descr\ Features can be moved in the same project as a result of refactoring, and also across projects, when developers incorporate it into another project. \movefeature{} combines two operators; \clonefeature{} (explained below) to clone the \featurecmd\ (and its mapped assets) to its new location, and \removefeature{} to remove it from its previous location.

\begin{comment}
\precond\ The source feature, its parent and the target parent are present in the given feature model; 

\looseness=-1
\postconds\ (1.) Feature is absent from the \subfeatures of its \textit{parent} feature. 
(2.) Source feature's \textit{parent} is set to the target feature.
(3.) Source feature is contained in \subfeatures of target feature.
\end{comment}

\noindent{}$\textbf{\makefeatureoptional}:\featurecmd\rightarrow\bool$ 

\descr\ Often, developers want to keep a feature's implementation in the $AT$, and decide whether to include it or not at compile time, instead of deleting it altogether. \makefeatureoptional{} sets a \featurecmd's \texttt{boolean} property \optional to \texttt{true}. By default, every \featurecmd\ is mandatory when added to the \featuremodel. This operator allows to keep the feature's implementation in the $AT$ while allowing developers to activate or deactivate the feature.

\noindent{}$\textbf{\clonefeature}:\,\featurecmd\times\featurecmd\rightarrow\bool$

\looseness=-1
\descr\ Cloning a \featurecmd\ manually requires developers to recollect its location in software assets. These assets can be of different types (directory, document, code artifact, text etc). 
Features can be scattered and therefore harder to locate. This is where the stored (and maintained) meta-data pays off. \clonefeature{} simply invokes a \textit{convenience} operator; \texttt{getMappedAssets}, to retrieve all assets mapped to the feature. It then clones the \featurecmd\ and all its mapped assets in the target $AT$ and $FM$. The operator also stores traces for the \asset\ and \featurecmd\ clones in the \textsl{TraceDatabase}. The \globalversion of the $FM$ is incremented and assigned to the target \featurecmd\ (\parent of the \featurecmd\ clone).

\begin{figure}[t]
	\vspace{-.1cm}
	\centering
	\includegraphics[width=.92\columnwidth]{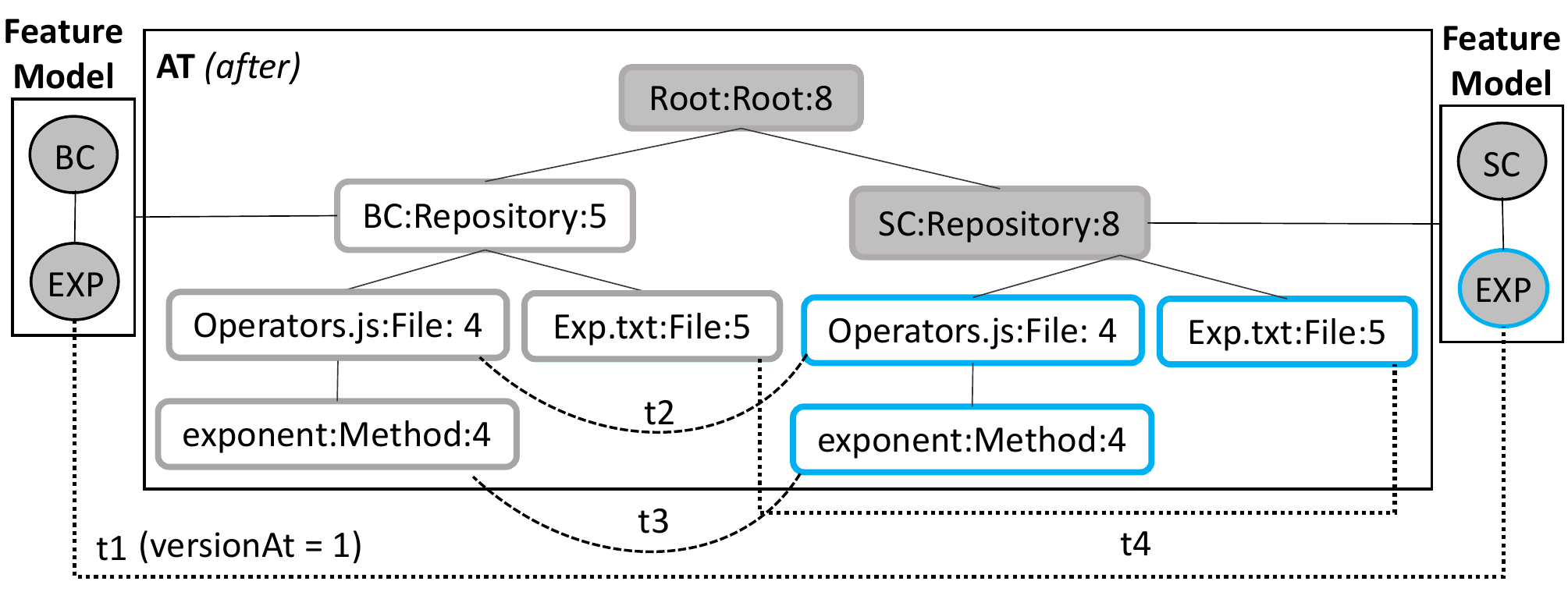}
	\vspace{-.3cm}
	\caption{Illustration of \clonefeature{(\textit{EXP}, \textit{SC})}}
	\label{fig:clonefeature}
	
\end{figure}

\begin{comment}
	\looseness=-1
	\postconds\ (1.) For the source feature and each of its sub-features, a clone exists in target feature's feature model.
	(2.) For each feature that was cloned, a trace exists in the \ftracedb, with the feature's version at the time of cloning.
	(3.) For any asset mapped to the source feature or one of its sub-features, a clone exists in the AT subtree associated to 
	(4.) For each asset that was cloned, a trace exists in the \tracedb, with the asset's version at the time of cloning. 
	
\end{comment}
\looseness=-1
\examp\ After adding the \featurecmd\ \textit{EXP} (using \addfeature), the developer added two assets in \featurecmd\ \textit{BC}, and later mapped them to \featurecmd\ \textit{EXP}. The assets are a method ``\textit{exponent}'' and a textual file ``\textit{exp.txt}'' with documentation of \textit{exponent}. The developer now wants to reuse \featurecmd\ \textit{EXP} in \textit{SC}. To clone the feature, she invokes \clonefeature{}, which clones the \featurecmd\ \textit{EXP} and its mapped assets to \textit{SC}. Additionally, traces for the \featurecmd\ and asset clones are added to the \textsl{TraceDatabase}. This example is illustrated in \Figref{fig:clonefeature}. Note that even though \textit{Operators.js} was not cloned, the virtual platform created a clone, as the method \textit{exponent} could not be added directly to the repository. This is referred to as \textit{tree slicing}, which the virtual platform adopts to ensure that the \wellformedness\ of the $AT$ is maintained.

\algo{
\setlength{\textfloatsep}{0pt}
\begin{algorithm}[h]
	\caption{Operator \clonefeature( SFP, TFP )}
	\label{alg:clonefeature}
	\fontsize{7}{7.5}\selectfont
	\begin{algorithmic}[1]
		
	\State SF = \traversepath(SFP)
	\State TF = \traversepath(TFP)
	\State SC =  SF.clone
	\State \addfeature(SC,TF)
	\For {$\textit{A} \leftarrow SF.\mappedassets$}
	\State slice = \getslice(A,SA)
	\State \addasset(slice,TA)
	\State Aclone = \getclone(slice)	
	\State \mapassettofeature(Aclone,SC))
	\EndFor
	\State \addtrace(SF,SC,SFM.version)
	\State \updateversion(TFM)
	\For{$\textit{S} \leftarrow SF.subfeatures$}
	\State \clonefeature(\getfeaturepath(S),\getfeaturepath(SC))
	\EndFor
	
\end{algorithmic}
\end{algorithm}
\setlength{\textfloatsep}{\textfloatsepsave}
}

\noindent{}$\textbf{\propagatefeature}:\featurecmd\times\featurecmd\rightarrow\bool$

\looseness=-1
\descr\ 
\propagatefeature replicates the changes in the \featurecmd\ (e.g., renaming, adding or removing sub-features) to either selected, or all of its clones. For checking if propagation is valid and necessary, it checks two conditions, based on the \textsl{TraceDatabase}. First, if one of the features provided is a clone of the other. Second, if the \featurecmd\ was modified after cloning (current \version\ > \textsl{versionAt}). After propagating changes, it creates new traces between the source and newly modified targets (both \featurecmd\ and \asset), and adds them to the \textsl{TraceDatabase}.

\begin{comment}
	\precond\ The target feature is a clone of the source feature.
	
	\looseness=-1
	\postconds\ (1.) All sub-feature added to the source after cloning are cloned in the target. 
	(2.) Any assets mapped to the feature or any sub-features after cloning have been cloned (by getting a slice of the mapped asset upto the asset having the feature model containing SF) and mapped. (3. ) The \globalversion of target feature model is updated. (4. ) The \globalversion of the AT is incremented after every \cloneasset, and assigned to all target assets during cloning. 

\algo{
\setlength{\textfloatsep}{0pt}
\begin{algorithm}[h]
	\caption{Operator \propagatefeature( SFP, TFP )}
	\label{alg:propagatefeature}
	\fontsize{7}{7.5}\selectfont
	\begin{algorithmic}[1]
		
	\State SF = \traversepath(SFP)
	\State TF = \traversepath(TFP)
    \If{\isclone(SF,TF)}
	\If{\detectchanges(SF)}
	\State \makeconsistent(SF,TF)
	\For{$\textit{A} \leftarrow SF.\mappedassets$}
	\If{!\assetcloned(A,TA)}
	\State slice = \getslice(A,SA)
	\State \addasset(slice,TA)
	\State Aclone = getCloneFromSlice(slice)	
	\State \mapassettofeature(Aclone,SF))
	\EndIf
	\EndFor
	\State \addtrace(SF,TF,SFM.version)
	\State \updateversion(TFM)
	\For{$\textit{S} \leftarrow SF.subfeatures$}
	\If{\featurecloned(S,TF)}
	\State SC = \getclone(S,TF)
	\State \propagatefeature(\getfeaturepath(S),\getfeaturepath(SC)
	\Else
	\State \clonefeature(\getfeaturepath(S),\getfeaturepath(TF))
	\EndIf
	\EndFor
	\EndIf
	\EndIf
\end{algorithmic}
\end{algorithm}
\setlength{\textfloatsep}{\textfloatsepsave}
}
\end{comment}

\noindent
\section{Prototyping and Evaluation}\label{sec:evaluation}
\looseness=-1
\noindent
We prototyped and evaluated the virtual platform qualitatively and quantitatively: (i) in a comparative assessment against the frameworks presented in \secref{sec:methodology}, (ii) using a simulation study based on revision histories from \co-based system.
All results to replicate the evaluation are in our appendix\,\cite{appendix:online}. 
\label{sec:implementation}
\looseness=-1
The prototype, implemented in Scala, provides an API as the main interface to execute the operators.
In the production-ready tool, this API would be usable as a command line interface or a set of IDE commands.
We used a strategic programming library (kiama) for efficient tree traversal and rewriting.
After implementing all operators, we created test scenarios to verify the correctness. These test scenarios were developed using domain knowledge acquired by experience, and also inspired by observing scenarios from the case study of Clafer Web Tools.
We checked correctness by comparing the result state (AT, trace, and mappings) after operator invocations to the expected one.
We also simulated the illustrative example presented in \secref{sect:calculatorexample} by automatically realizing all the discussed scenarios.

\vspace{-0.1cm}
\noindent
\longv{\subsection{ExpressionSolver Simulation}
We introduced the illustrative example \textit{ExpressionSolver} in \secref{sect:expressionsolver}, where we used feature cloning to reuse the functionality from \textit{BasicCalculator} and \textit{Stack} into \textit{ExpressionSolver}. We now discuss the realization of the three sub-systems, and the simulation of scenarios presented in \secref{sect:expressionsolver}.

We implemented the three sub-systems (\textit{BasicCalculator}, \textit{Stack} and \textit{ExpressionSolver}) using JavaScript. Every sub-system was contained in one file, where methods contained the implementation for all features. For example \textit{BasicCalculator} consisted of four methods; \textins{add}, \textit{subtract}, \textit{multiply} and \textit{divide}. Using the \textit{JavaScript Parser}, we loaded the files along with their methods in the AT. Then, we created one feature model per file, and assigned it to the file assets in the AT (\textit{BasicCalculator}, \textit{Stack} and \textit{ExpressionSolver}). Using \mapassettofeature, we mapped every asset to it's respective feature. For example, \textit{PUSH} feature was mapped to the \textit{push} method in \textit{Stack}. While mapping, the features were added to the feature model of the respective file on-the-fly. At this point, the AT contains all assets mapped to their corresponding features in their feature models. 

We then cloned the features; \textit{ADD}, \textit{SUB} and \textit{MULT} from \textit{BasicCalculator}, and \textit{PUSH} and \textit{POP} from \textit{Stack} into \textit{ExpressionSolver}. Cloning of features resulted in cloning of their mapped assets into the sub-AT of \textit{ExpressionSolver} as well. After the initial setup and feature cloning, the \tracedb and \ftracedb hold traces for all assets and features involved respectively.
Figure \ref{fig:basiccalculator1} shows the three systems before the simulation of the scenarios explained in \secref{sect:expressionsolver}. 

Now we discuss how Virtual Platform smoothly manages the scenarios discussed in \secref{sect:expressionsolver}.

\myparhead{Scenario 1 (Clone a feature)}: In the first scenario, the \textit{DIV} feature was cloned from \textit{BasicCalculator} to \textit{ExpressionSolver}. When the feature was cloned, a trace between \textit{DIV} feature and its clone was created and added to the \ftracedb. Also, the mapped assets of \textit{DIV} feature (in this case, the \textit{div} method) were cloned in \textit{ExpressionSolver.js} automatically, and an asset clone was added for the \textit{div} method and its clone in the \tracedb. Adding these traces allows retrieving this information later for feature location and change propagation. Figure \ref{fig:basiccalculator2} shows the simulation of the first scenario, where \textit{DIV} feature is cloned from \textit{BasicCalculator} into \textit{ExpressionSolver}. 

\myparhead{Scenario 2 (Propagate a bugfix)}: In the second scenario, the implementation of the \textit{div} feature changed to incorporate a check for division by zero. After making this change, we needed to propagate this to the places where the feature is cloned. The change was incorporated as a \changeasset, which updated the \version of the related assets (\textit{divide} method). We then used \propagateasset, which compared the \version of the \textit{divide} method in \textit{BasicCalculator} to the \version at which the asset was cloned. After determining that a propagation is in fact due, both versions were made consistent. This was achieved by adding one line for \propagateasset. Clone detection was not required during change propagation, as \tracedb stored all traces during initial cloning. After propagating changes from the \textit{div} method to its clone, a new trace was added in the \tracedb, signifying that a change propagation occurred, and that both versions are up to date and consistent with each other up till the version of \textit{div} method in \textit{BasicCalculator.js} (\texttt{versionat}). This scenario is illustrated in Figure \ref{fig:basiccalculator3}, where a change in \textit{DIV's} implementation (represented by a delta sign) is propagated to it's clone in \textit{ExpressionSolver}.

\myparhead{Scenario 3 (Propagate a new feature)}: In the third scenario, \textit{BasicCalculator} was modified to add the implementation of exponent feature (EXP). The feature was added using \addfeature, and the method corresponding to the feature was added using \addasset. The mapping was created using \mapassettofeature. Please note that this is the simulation, whereas in real life, the asset-oriented operators like \addasset and \mapassettofeature will involve zero overhead, and would be invoked by the Virtual Platform. The only additional effort will be that for adding feature annotations. Once the newly added feature and asset were in place, and mapped to each other, we simply used \clonefeature which copied and added the feature EXP along with its implementation to \textit{ExpressionSolver}. In addition to cloning, \clonefeature created a mapping between the feature clone and the asset clone in the same way as they were linked in the \textit{BasicCalculator} (using \mapassettofeature), and added two traces; one in \ftracedb (between \textit{EXP} feature and its clone) and the other in \tracedb (between \textit{exponent} method and its clone). 

\begin{figure}[t]
	\vspace{-.4cm}
	\centering
	\includegraphics[width=\columnwidth]{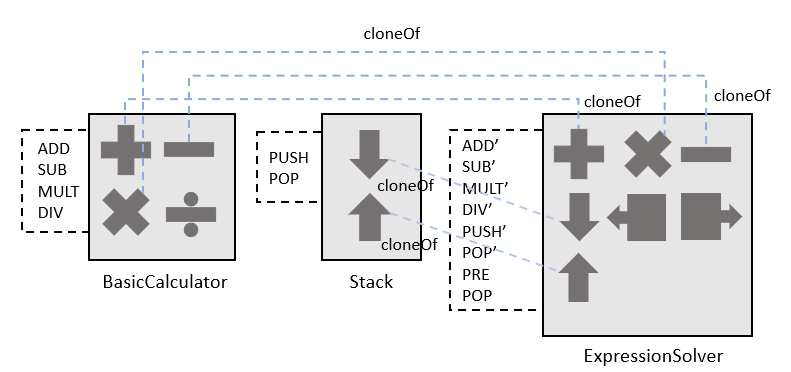}
	\vspace{-.4cm}
	\caption{Recording provenance data during feature cloning}
	\label{fig:basiccalculator1}
\end{figure}

\begin{figure}[t]
	\vspace{-.4cm}
	\centering
	\includegraphics[width=\columnwidth]{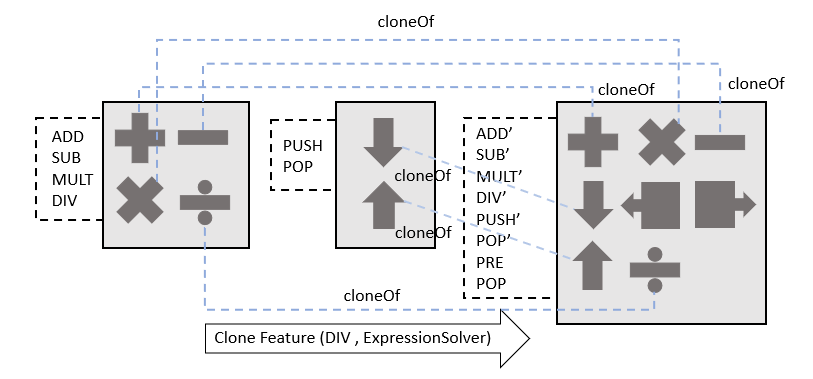}
	\vspace{-.4cm}
	\caption{Scenario 1: Cloning DIV feature in ExpressionSolver}
	\label{fig:basiccalculator2}
\end{figure}

\begin{figure}[t]
	\vspace{-.4cm}
	\centering
	\includegraphics[width=\columnwidth]{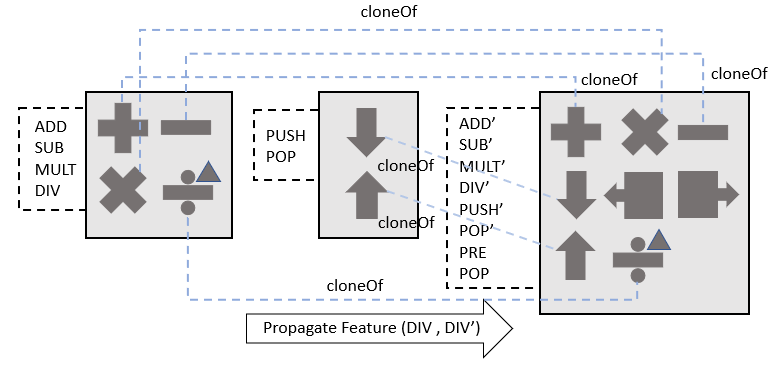}
	\vspace{-.4cm}
	\caption{Scenario 2: Propagating bug fix in ExpressionSolver}
	\label{fig:basiccalculator3}
\end{figure}

\begin{figure}[b]
	\vspace{-.4cm}
	\centering
	\includegraphics[width=1.05\columnwidth]{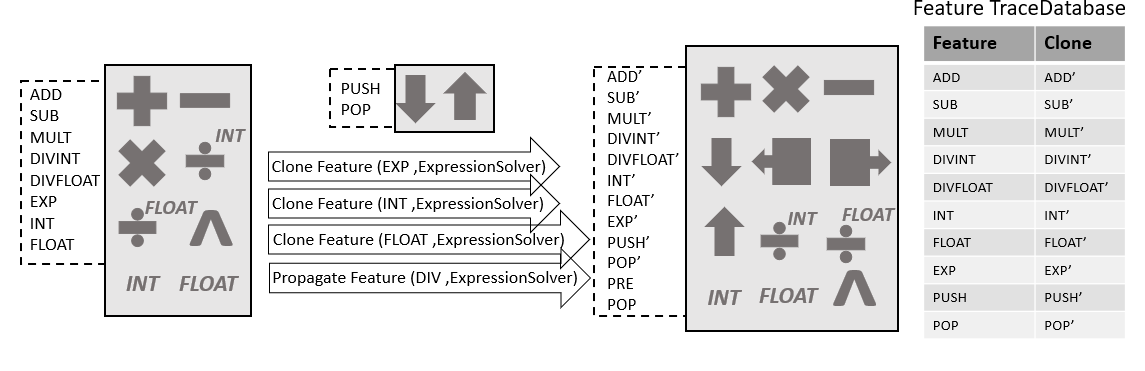}
	\vspace{-.4cm}
	\caption{Scenario 3 \& 4: Cloning \textit{EXP} feature and propagating \textit{NUMTYPE} feature}
	\label{fig:basiccalculator4}
\end{figure}
\myparhead{Scenario 4 (Propagate a combined feature addition and change to an existing feature implementation)}: In the fourth scenario, the \textit{BasicCalculator} was enhanced to include the ability to handle two data types. The feature \textit{NumType} was added to incorporate two data types; Integer (\textit{INT}) and Float (\textit{FLOAT}). The functionality of the \textit{divide} method was also diverged into two implementations, depending on the input type. This resulted in changes in asset and feature model versions. For propagating the changes, we simply called the operator \propagatefeature. The operator not only added the feature \textit{NumType} with its \subfeatures, it also cloned the assets mapped to these features such that the changes in the \textit{divide} method were also propagated to \textit{ExpressionSolver}. After cloning, feature traces for \textit{NumType}, \textit{INT} and \textit{FLOAT} were added to the \ftracedb. The resulting change propagation for the \textit{divide} method also lead to the addition of an asset trace in the \tracedb. Figure \ref{fig:basiccalculator4} illustrates the simulation of scenario 3 and 4, which involves cloning of three features (\textit{EXP},\textit{INT} and \textit{FLOAT}) from \textit{BasicCalculator} to \textit{ExpressionSolver}, and propagation of changes in the \textit{DIV} feature in \textit{ExpressionSolver}.
\myparhead{Scenario 5 (Make feature optional)}: In the fifth scenario, the \textit{EXP} feature was modified to be an optional feature. This was performed by invoking the \makefeatureoptional operator, which simply changed the feature's \texttt{optional} parameter to true. Note that this added operator allows storing an additional property of the feature and doesn't require the developer to know the code or locate a particular feature to change it's selectivity. Also, this step requires zero overhead in terms of time and memory, since it only changes an attribute of the feature, and doesn't require saving asset or feature traces for later use.  
\myparhead{Scenario 6 (Integrate cloned variants)}: Finally, since the \textit{BasicCalculator}, \textit{Stack}, and \textit{ExpressionSolver} can work as stand-alone systems, these sub-systems can be integrated to form a product line. This requires a variant comparison such that multiple implementations can me merged into a consolidated product line. \wm{Need Thorsten's view point on this}
\myparhead{Qualitative Cost and Benefit Analysis}: In the small illustrative example, we were able to simulate the operators to provide full reuse of both assets and features in \textit{BasicCalculator} and \textit{Stack}. During cloning, any costs for feature location costs were saved, and in propagation, costs for clone detection were saved. The only added cost arose when applying feature-oriented operators.
We argue that, feature annotations, if they are added by the developers during implementation, may enable a more efficient and accurate process than a one-time retroactive feature location step.
\ds{What about the cost for entering parameters of feature-oriented operators (see next section)? The more general question: What precisely are the ``items'' whose cost we measure (both when using VP vs. when using the traditional setup)?}
 
	\begin{table*}[h]
		\caption{Cost analysis of feature-oriented operators, with number of invocations (Invocs.) and cost per operator.}
		\label{tab:costandbenefit}
\begin{tabular}{ll|ll|ll|ll|ll|l}
\toprule
\textbf{Feature-oriented Operator}                               & \textbf{Cost} & \multicolumn{2}{l|}{\textbf{ClaferMooVisualizer}} & \multicolumn{2}{l|}{\textbf{ClaferIDE}} & \multicolumn{2}{l|}{\textbf{ClaferConfigurator}} & \multicolumn{2}{l|}{\textbf{ClaferToolsUICP.}} & \textbf{Total} \\

                                       &            & Invocs.            & Cost           & Invocs.       & Cost      & Invocs.           & Cost           & Invocs.                & Cost               &   Cost    \\
\hline
\Tstrut \addfeaturemodeltoasset & 2          & 1                      & 2              & 1                 & 2         & 1                     & 2              & 1                          & 2                  & 8     \\
\addfeature             & 2          & 66                     & 132            & 43                & 86        & 49                    & 98             & 51                         & 102                & 418   \\
\mapassettofeature      & 2          & 246                    & 492            & 137               & 274       & 117                   & 234            & 149                        & 298                & 1298  \\
\makefeatureoptional    & 1          & 0                      & 0              & 0                 & 0         & 0                     & 0              & 0                          & 0                  & 0     \\
\movefeature             & 3          & 16                     & 48             & 1                 & 3         & 4                     & 12              & 0                          & 0                  & 63    \\
\removefeature          & 1          & 27                     & 27             & 10                & 10        & 3                     & 3              & 0                          & 0                  & 40    \\
\clonefeature           & 2          & 20                     & 40             & 5                 & 10        & 1                     & 2              & 0                          & 0                  & 52    \\
\propagatefeature       & 2          & 0                      & 0              & 5                 & 10        & 4                     & 8              & 0                          & 0                  & 18   \\
\bottomrule
\end{tabular}
\end{table*}

}

\subsection{Comparative Evaluation}\label{sec:compeval}
\looseness=-1
\noindent
For comparison, we extracted activities supported by techniques for supporting \co (a.k.a., clone management), or the migration of cloned variants to an integrated platform (a.k.a., product-line migration). In total, we extracted 12 activities which we found to be common across most, if not all, existing techniques. We evaluated the virtual platform's ability to support the scenario from \secref{sec:motivation} and the 12 activities of related frameworks. Details are in the appendix\,\cite{appendix:online}. 

\tabref{tab:smallcomparison} shows whether and how an activity related to either clone management or product-line migration is supported by an existing framework, as well as virtual platform. The activities are: feature identification (features defined in a variant), feature location (recovering traceability between features and assets), feature dependency management (managing constraints among features), feature model creation (creating and evolving a feature model), storing feature-to-asset mappings, clone detection (identifying assets which are clones of one another), feature cloning (ability to clone features), change propagation (replicating changes made in an asset to its clone), creation of reusable assets (which can be used to derive variants), product derivation (ability to derive a partial or complete product given a configuration), variant integration (merging assets/variants by taking variability into account), and variant comparison (comparison of assets to find commonalities and varaibilities). 

In summary,
\begin{table}[t]
	\caption{Comparison of the virtual platform with activities supported by clone-management and product-line migration frameworks}   
	\vspace{-.3cm}

		\label{tab:smallcomparison}
		\begin{tabularx}{\linewidth}{p{8.6cm}}
			\toprule
			\textbf{Feature identification }$\rightarrow$  abstract operator\,\cite{Rubin2015}, specified in the beginning\,\cite{fischer.ea:2014:ecco,variantsync:2016,MartinezBut4Reuse}, specified any time in virtual platform  \\
			
			\textbf{Feature location} $\rightarrow$ abstract operator\,\cite{Rubin2015}, extracted \cite{fischer.ea:2014:ecco,MartinezBut4Reuse}, internal tagging\,\cite{variantsync:2016}, also internal tagging in virtual platform \\
			
			\textbf{Feature dependency management} $\rightarrow$ abstract operator\,\cite{Rubin2015}, statically mined\,\cite{MartinezBut4Reuse}, specified in beginning\,\cite{variantsync:2016}, specified any time in virtual platform\\
			
			\textbf{Feature model creation} $\rightarrow$ multiple abstract operators\,\cite{Rubin2015}, activity \cite{MartinezBut4Reuse}, specified in the beginning\,\cite{variantsync:2016}, dynamically grows in virtual platform   \\
			
			\textbf{Feature-to-asset mapping} $\rightarrow$ abstract operator\,\cite{Rubin2015}, extracted\,\cite{fischer.ea:2014:ecco,MartinezBut4Reuse}, specified any time\,\cite{variantsync:2016}, specified any time in virtual platform    \\
			
			\textbf{Clone detection} $\rightarrow$ textual diff tools\,\cite{Rubin2015}, feature expression comparison\,\cite{variantsync:2016}, git clone points to source\,\cite{montalvillo2015tuning}, not needed in virtual platform   \\
			
			\textbf{Feature cloning} $\rightarrow$ supported by virtual platform  \\
			
			\textbf{Change propagation} $\rightarrow$  multiple abstract operators\,\cite{Rubin2015}, variant synchronization\,\cite{variantsync:2016}, using Git merge\,\cite{montalvillo2015tuning}, automated in virtual platform\\
			
			\textbf{Reusable assets creation} $\rightarrow$ abstract \& incremental\,\cite{Rubin2015}, reuse existing \\variants \,\cite{fischer.ea:2014:ecco}, reusable core assets\cite{MartinezBut4Reuse,montalvillo2015tuning} and features in virtual platform\\
			
			\textbf{Product derivation} $\rightarrow$  abstract\,\cite{Rubin2015}, customizing after cherry-picking\,\cite{montalvillo2015tuning}, composition\,\cite{fischer.ea:2014:ecco,MartinezBut4Reuse}, preprocessor-like in virtual platform \\
			
			\textbf{Integration} $\rightarrow$ abstract operator using meta-data\,\cite{Rubin2015}, third party tool\,\cite{variantsync:2016}, Git merge\,\cite{montalvillo2015tuning}, \edit{manual or tool-based, guided by meta-data in the virtual platform} \\
			
			\textbf{Variant synchronization} $\rightarrow$ Git dif\,\cite{montalvillo2015tuning}, code comparison\,\cite{fischer.ea:2014:ecco,MartinezBut4Reuse}, not needed in virtual platform \\
			\bottomrule
		\end{tabularx}

\end{table}
among all frameworks, the virtual platform is the first one fully committed  to recording traceability, instead of recovering it later. 
It automatically maintains traces between cloned assets, and encourages developers to map features to assets of all types and all granularity levels (not just code blocks).
This traceability has a cost to developers; however, at the same time, it can significantly reduce cost when complex evolution activities are performed, as detailed below.

\looseness=-1
The other frameworks define their involved activities either abstractly or using heuristics (e.g., feature location).
The virtual platform includes exact specifications and implementations of opera\-tors---possible since we address a broad range of evolution scenarios, rather than just the ``big bang'' scenario of platform migration.
The existing methods have not been applied to real project revision histories as part of their evaluation, rather explain that they support migration scenarios described before.

\subsection{Simulation Study}
\looseness=-1
\noindent
We used an open-source system called Clafer Web Tools (CWT, \cite{antkiewicz2013clafer}) that was evolved using \co in three cloned variants (\textit{ClaferMooVisualizer}, \textit{ClaferConfigurator}, \textit{ClaferIDE}) towards an integrated platform (\textit{ClaferUICommonPlatform}), including many feature clonings across the variants.
We evaluated the virtual platform's efficiency by simulating the evolution of CWT, retrofitting our operators to achieve the original evolution, and studying the costs and benefits.

\longv{Specifically, our traditional operators are cost-neutral, as they represent the operations typically used in IDEs and version control systems.
Our feature-oriented operators have a cost, as they require the developer to specify parameters, such as features and feature locations.
Their prospective benefit arises as they avoid a need for feature location and clone detection.}

\longv{The four considered sub-systems of CWS were developed in a timeframe between October 2012 and February 2014.
\textit{ClaferMooVisualizer} creates optimized model instances and visualizes them. \textit{ClaferIDE} supports development using Clafer, and allows developers to write, compile and run models. \textit{ClaferConfigurator} is used for constraint generation based on desired instances. \textit{ClaferUICommonPlatform} integrates all functionality from all three sub-systems mentioned above. 
The implementation language of all sub-systems is JavaScript. 
\textit{ClaferMooVisualizer} was the first to be developed, followed by \textit{ClaferConfigurator}, which includes a considerable proportion of functionality cloned from \textit{ClaferMooVisualizer}. The third project \textit{ClaferIDE} was developed by cloning functionality from both existing projects, and lastly, \textit{ClaferUICommonPlatform} was created by cloning big chunks of functionality from all three former projects.  
There are common features shared by some tools, and unique ones distinguishing individual tools from one another.}

\newcommand{\paperonly}[1]{#1}
\newcommand{\appendixonly}[1]{}
\appendixonly{\subsection{Additional details for Clafer Web Tools case study (Section~\ref{sec:evaluation})}}

\appendixonly{\subsubsection{Used Dataset}}
 We used a dataset by Ji et al.\,\cite{ji2015maintaining} that augments the original codebase with feature information, as if it had been developed in a feature-oriented way.
It comprises a full revision history for the four sub-systems, with source code from the original developers, and feature information manually added by researchers.
Feature information is contained in three types of artifacts: feature models, feature-to-asset mapping files, and embedded feature annotations in code.
\appendixonly{To create this information, Ji \textit{et al.} consulted with the original developers to map files and folders from the repositories to various features, and added annotations in source code. They attempted to reconstruct the cloning of features between repositories and add appropriate meta-data. 
The log for cloning activity during reconstruction was kept in a dedicated spreadsheet, which consisted of details including the feature that was cloned, the source repository from where the feature was cloned, and the target repository. It also included the commit hashes of the source and target repositories.

After introducing the feature-oriented changes, they committed to the online repository after merging the changes.}
\paperonly{We provide details about the dataset in our appendix\,\cite{appendix:online}.} \appendixonly{Feature Models were provided in Clafer format \cite{bak.ea:2010:sle}, that is, as text files with the file extension ".vp-project". 
Each such file contains a list of features,
 
one per line, with indentation to define the hierarchy. Every feature is indented one tab more than its parent feature. The relationship between features (Or, And, Xor) is stated in the first child feature; the name of the feature follows the keyword determining the relationship. Feature names are unique per feature model.}

\appendixonly{Feature-to-asset mappings are given in two different ways, based on the granularity of the mapping:
Mappings to folders and files are specified in separate  files, containing a list of the names of the features the folder or file is mapped to.
Mappings to segments of a file are specified by inline annotations in the file, specifying the start and end of a code block.

A code block consists of consecutive lines of code, beginning with \textit{ "//~\&begin[feature name]"} and ending with \textit{"//~\&end[feature name]"}. One block can be mapped to multiple features, and one feature can be implemented via multiple blocks. To map a block to multiple features, a comma-separated list of feature names can be given in the square brackets. Line annotations are a special case of code annotations, where the block is a single line of code. Line annotations have a comment \textit{"//~\&line[feature name]"} at the end of the line.}

\looseness=-1
\parhead{Performing the Simulation.}
We retrofitted CWT's full revision history to our operators to extract a sequence of (high-level) operator applications that accurately capture the changes previously expressed by the history of (low-level) file-based commits.
We analyzed each pair of successive commits to extract a set of operator applications that produces the delta between the commits.
Replaying the operator applications in the given order creates and updates the AT.

\appendixonly{For our simulation, we required a list of all commits of all four subsystems. By running a "\textit{git log}", we retrieved the commits along with their time stamp. The commits for all subsystems were then sorted with respect to their time stamp to get an order for the simulation. These were used as the input for our simulation. }

\appendixonly{Based on the (low-level) change set, we derived a set of (high-level) operator applications performed in the commit in a semi-automated process.

 We first used diff with "-\,-name-status", which produced a list of files, along with the operation that was performed on them after the last commit. The first character of every line  specified the change, followed by a tab and the file name. The first letter "A" represented addition, "M" modification, "D" deletion, and "R" rename or move.
Based on their file extensions, we distinguished files as feature files or artifact files. We observed that we need to process the list of files in a certain order, for the following rationale: Executing the operations in the lexical order of file names leads to dependency-related problems. For example, \mapassettofeature cannot be applied if the asset to be mapped is only added later as part of the same commit. Therefore, we set a precedence, which is explained below:

\begin{enumerate}	
		\item \textbf{Feature model modification}: To enable those operators that rely on available features (e.g., feature cloning and propagation), we first need to sync the feature models in the asset tree with the ones from the repository. For that, we read the ".vp-project" file line by line, and added all features not present in the feature model. Also, we deleted the features present in the feature model if they were not in the latest \version of the feature model file. Lastly, we moved features that were already in the feature model, but moved to another place in the hierarchy. The operators relevant here were \addfeature, \removefeature, \movefeature and \addfeaturemodeltoasset. Another possible type of feature model modification, feature model deletion, did not arise in the case study and did not have to be addressed.	
		
		\item \textbf{Asset tree modification}: Addressing those operations which rely on an available artifact, we ensured that the artifact tree is in sync with the related changes from the commit. Newly added assets from the commit were added as nodes to the asset tree, with their asset type being either set to \textit{folder} or \textit{file}. If a code file was added, we parsed the file using \textit{ParseCodeFile} to annotate the functions and blocks with annotations added by the developer. For modifications in code files, we removed the sub-artifacts of the file artifact, and used \textit{ParseCodeFile} to parse and load the sub-artifacts again. For deletion of artifacts, we unmapped the artifacts to their mapped features and deleted the artifacts from the asset tree.	
		
		\item \textbf{Folder and file mapping}: With the feature models and artifact tree being in sync, we considered changes related to the mapping of folders and files, as specified in newly added, modified, or deleted files with the extension .vp-folder and .vp-files. We updated the mappings in our AT accordingly. For every new mapping, we checked if both the artifact and the feature were in the AT and FM respectively, and mapped the asset to the feature.

		\item \textbf{Clone Feature}: Finally, we investigated if any features were cloned during the commit. We used Ji \textit{et al.}'s spreadsheet to mine feature clones, and create traceability links between those features and their clones. We also used the information to create feature-to-artifact mappings in the target repository by retrieving the mapped artifacts of the source feature and finding their clones in the target repository. This resulted in the addition of artifact traces in \tracedb, and feature traces in \ftracedb.

		We needed to make sure that the mapped artifacts of the feature in source and target repositories are same. To this end, we looked at the mapped assets manually, but it is important to note that this is also an automatable task, we however had to look manually as we were reverse engineering from the case study instead of using the Virtual Platform in parallel to development. 37 out of 48 artifacts in source and target were similar (clones of each other). 
		
		\item \textbf{Propagate Feature}: As mentioned earlier, Ji \textit{et al.} maintained a separate file for recording the cloning activities during the actual reconstruction. The file doesn't contain information about any feature propagations. We therefore decided to manually assess the artifacts to look for any change propagations that might have occured. To this end, we check for the following conditions:
		\begin{enumerate}
			\item The artifact \textit{A} is mapped to a feature \textit{F}.
			\item \textit{F} is cloned from a source repository \textit{S}.
			\item \textit{A} is cloned from the AT of \textit{S}.
			\item \textit{A}'s \version is ahead of it's source artifact in \textit{S}.
		\end{enumerate}

		In simple words, if the artifact was cloned as part of a feature cloning, and the original artifact is modified after the cloning, then it is possible that the same changes are (manually) performed by the developer in the target artifact. One of the authors manually inspected such cases, and found some overlapping and some non-overlapping changes. This can be explained by the fact that the developers might have cloned the basic functionality, and later enhanced the versions independent of each other. It is important to note that cloning is still useful, as it helps to eliminate the redundant  efforts by developers, and keeps a track of changes being made to the versions in parallel to each other.

\end{enumerate}}
	
\newcommand{\Cfeatform}{C_{\textit{\small feat}}}
\newcommand{\Cfeatopform}{C_{\textit{\small feat}}(op)}
\newcommand{\Cmissedform}{C_{\textit{\small miss}}}
\newcommand{\Clocform}{C_{\textit{\small loc}}}
\newcommand{\Ccloneform}{C_{\textit{\small clone}}}
\newcommand{\Cfeat}{$\Cfeatform$}
\newcommand{\Cfeatop}{$\Cfeatopform$}
\newcommand{\Cmissed}{$\Cmissedform$}
\newcommand{\Cloc}{$\Clocform$}
\newcommand{\Cclone}{$\Ccloneform$}

\newcommand{\todo}[1]{}

\looseness=-1
\parhead{Cost \& Benefit.} \label{sec:costandbenefit}
As costs, we measure the additional effort imposed on developers by our platform.
Our traditional, asset-oriented operators (left-hand column of \tabref{tab:operatorsusagetable}) do not lead to additional cost, because these tasks are performed in traditional development as well.
Cost arises from two components, both related to our feature-oriented operators (right-hand column of \tabref{tab:operatorsusagetable}): one called \Cfeat\ for maintaining features, one called \Cmissed\ for dealing with omissions during feature maintenance. The latter arises if the developer forgets to invoke a feature-oriented operator and then later the feature information is missing for a relevant feature-oriented activity.

As benefits, we consider the saved cost in two dimensions: feature location and clone detection.
Feature location cost \Cloc\ is saved on invocations of certain operators that rely on previously specified mappings.
Clone detection cost \Cclone\ is saved on invocations of one certain operator for propagating changes along clones from our clone database.

We study these costs and benefits in four dedicated research questions.
RQ1 and RQ2 are devoted to costs, while RQ3 and RQ4 are devoted to benefits.
We first discuss these research questions, before weighing off the observed costs and benefits.

\smallskip
\looseness=-1
\parhead{RQ1.}  \textit{What are the costs of maintaining features using feature-oriented operators?}
The overall cost \Cfeat\ arises from accumulating the cost of applying feature-oriented operators.
Each feature-oriented operator $op$ has a cost \Cfeatop$\,=\, \#invoc(op) * cost_{abs}(op)$, which depends on the number of invocations of $op$, and the absolute cost of each invocation of $op$.
Based on \tabref{tab:operatorsusagetable}, there are 724  invocations of feature-oriented operators in total. 
Two operators contribute the bulk to this number, namely \texttt{MapAssetToFeature}~(368) and \texttt{AddFeature}~(229).
The absolute cost per invocation can be assumed to be low (in the order of seconds) because it mostly amounts to picking the feature name, when it is fresh in the developer's mind.
An exception are situations where the developer has to deal with earlier omissions (see RQ2).

\smallskip
\looseness=-1
\parhead{RQ2.} \textit{What percentage of feature maintenance operations required additional feature location effort?}
The omission-related cost \Cmissed\ arises from the number of late invocations of \texttt{MapAssetToFeature}, representing situations where the developer missed to specify an asset-to-feature mapping when the asset was added.
This number is to be multiplied by the absolute cost for these invocations, which is generally higher than a regular invocation.
Our operators
\texttt{CloneFeature}, 
and 
\texttt{PropagateToFeature} rely on a complete mapping from a feature to its assets.
A third relevant operator is \texttt{AddFeature} which adds feature information to source code added earlier.
In absence of a recorded mapping, each operator requires an expensive manual feature location step, which is not required in our approach (see RQ3).
We counted the number mappings that were added before or after one of these operators was invoked, which indicates that the researcher preparing the original dataset noticed an omission.
We determined 14 relevant mappings for \texttt{CloneFeature} (2 relevant invocations, 3.7\% overall).
and 25 relevant mappings for \texttt{AddFeature} (12 relevant invocations, 4.0\% overall).
We did not discover any relevant mappings for \texttt{PropagateToFeature}, yielding 39 late invocations in total.

\begin{table}[t]
\caption{Operator invocations in simulation study:asset-oriented and feature-oriented operators}
\vspace{-.3cm}
\begin{adjustbox}{max width=1\linewidth}
\label{tab:operatorsusagetable}
\begin{tabular}{llllll}
\toprule
\textsf{operator}      & \textsf{freq.} & \textsf{operator}   & \textsf{freq.}\\
\midrule
AddAsset                & 3,527  & AddFeature             & 229       \\
ChangeAsset             &1,191   & AddFeatureModelToAsset  &4         \\
RemoveAsset            &1,060   & MapAssetToFeature       &368       \\
MoveAsset             &303       & RemoveFeature           & 40        \\
CloneAsset             &48      & MoveFeature             & 22        \\
PropagateToAsset          &8     & CloneFeature           &	54        \\
 &                                & PropagateToFeature        & 7         \\
\bottomrule\\
\end{tabular}
\end{adjustbox}
\end{table}

\smallskip
\looseness=-1
\parhead{RQ3.}  \textit{To what extent can feature location costs be avoided when using feature-oriented operators?}
The operators \texttt{CloneFeature}  and 
\texttt{Propagate\-Feature} rely on previously specified mappings.
Conversely to RQ2, we can assume that each invocation of one of these operators avoided manual feature location when it did not require any fixing of omitted annotations.
So, we define \Cloc{} to rely on the number of feature location steps saved by an invocation of one of our operators.
We count 54 invocations of \texttt{CloneFeature}, and 7 relevant invocations of \texttt{PropagateToFeature}, leading to a final value of 61.
This number is to be multiplied with the absolute cost of feature location, which can be assumed to be high (earlier work \cite{ji2015maintaining} gives an estimate of 15 minutes per feature), based a strong reliance on the developers' memory, and an understanding of how cross-cutting features are scattered.

\smallskip
\looseness=-1
\parhead{RQ4.}  \textit{To what extent can clone detection costs be avoided when using feature-oriented operators?}
Since the propagation of changes along clones requires a complete specification of the clones at hand, we can assume that every application of \texttt{PropagateToFeature} saves one application of clone detection (either manual or using a tool).
In our subject system, we identified 7 invocations of \texttt{PropagateToFeature}.
To obtain the value of \Cclone, this number of is to be multiple with the absolute cost for clone detection.
Manual clone detection is a tedious and error-prone task, and known to be infeasible for larger systems \cite{koschke2007survey}.
Tool-based clone detection requires manual verification and postprocessing, since even the most advanced clone detection tools have imperfect precision and recall \cite{sajnani2016sourcerercc}.

\subsection{Discussion}

\edit{\looseness=-1
\parhead{Break-Even Point.}
We can now weigh off the costs observed in RQ1+2 against the benefits from RQ3+4.
Consider the following formula, which specifies the total benefit of using the virtual platform:  $B_{total}$ = -(\Cfeat\ + \Cmissed) + (\Cloc\ + \Cclone).
If this formula yields a positive value, the virtual platform surpasses the break-even point and leads to a net benefit.

The value of $B_{total}$ depends on the absolute costs for operator invocations, feature location, and clone detection, which are unavailable.
However, we can perform an approximation based on plausible estimates:
(1.) For the cost of feature location, we rely on the earlier literature estimate \cite{ji2015maintaining} of 15 minutes per instance.
(2.) We assume clone detection to have the same cost as feature location.
(3.) We assume the cost for adding an omitted annotation to be 10 times as high as a regular operator invocation.
Based on these three assumptions, we break even if \textit{invoking a feature-oriented operator takes 54 seconds or less} on average.
In practice, the benefit can be assumed to be  larger, since invoking a feature-oriented operator mostly entails picking a feature name (while the feature is still fresh in the developer's mind), a matter of a few seconds.}

\edit{This calculation shows promising results in terms of saved effort and time. 
By simulating the development of the case study with feature-oriented information, 
we can reuse as much as 20 features from one project (\textit{ClaferMooVisualizer}) by cloning them. We envision greater accuracy and efficiency levels when the virtual platform is used alongside development.}

\looseness=-1
\parhead{Representativeness.}
Our case is representative for systems of comparable size (547k lines, four variants). Many reported product-line migrations are of similar size \cite{martinez2017espla}. We argue for representativeness for larger systems qualitatively. Our case has all evolution activities observable in industrial systems, supported by other frameworks. Still, the virtual platform is evaluated more extensively than any of these.

%\edit{\parhead{Representativeness.} Our case represents most systems of comparable size (i.e., 54.7K lines of text) and number of variants (four). Many product-line migrations of comparable-sized cases are reported in the catalogue in\,\citep{martinez2017espla}. For larger systems, the representativeness still holds, as our case undergoes all evolution activities typical in most software systems, which are also supported by other frameworks (detailed comparison in\,\cite{appendix:online}). Still, the virtual platform is evaluated much more thoroughly.}

\parhead{Threats to Validity.}
A threat to external validity is that our operators do not completely capture the real-world scenarios developers encounter when dealing with variant-rich systems. We mitigate this threat with our evaluation based on the simulation of a real system.
There is a general lack of available systems for benchmarking on realistic revision histories with available feature information, a problem that we aim to address as part of our ongoing benchmarking initiative \cite{struber2019facing,berger2019dagstuhl}.

\looseness=-1
There are two main threats to internal validity.
First, our calculation of \Cmissed\ could be incomplete: there might be potential omissions not fixed by a later commit.
This situation is comparable to other research that relies on potentially imperfect datasets (e.g., in software defect prediction \cite{catal2009systematic,struder2020feature}). 
While our analysis focuses on omissions that later required fixing, these omissions are arguably the most relevant ones in practice.
Second, there could be implementation errors; after retrofitting our operations to the development process given by the commit revision, the $AT$ might be in an incorrect state. To mitigate this threat, one author, not involved in the simulation, manually inspected a random sample of 25 commits by comparing the \textsf{git} \textit{diff} with the $AT$ resulting from operator invocations. The $AT$ was always consistent.

\section{Related Work}\label{sec:related}
\looseness=-1
\noindent
The five most closely related works are the clone-management and product-line-migration frameworks that we used to inform the virtual platform's design (\cite{Rubin2015,fischer.ea:2014:ecco,MartinezBut4Reuse,variantsync:2016,montalvillo2015tuning}, cf. \secref{sec:methodology}).
In \secref{sec:compeval} and our online appendix \cite{appendix:online}, we provide a detailed comparison, highlighting unique benefits of the virtual platform:
support for early traceability recording, operators for the full spectrum between the extremes ad hoc \co and  integrated platform, and an evaluation on a real project revision history.
We now discuss further related work on product-line migration and integrated-platform evolution.

\looseness=-1
The idea of automatically handling variation points, as the virtual platform does, is not new. In fact, going back to the 1970s, researchers have built so-called variation-control systems\,\cite{Linsbauer2017,linsbauer2020varcs}, which never made it into the practice of software engineering. These systems have been realized upon different back- and frontends (e.g., version-control systems\,\cite{munch.ea:1993:uniformversioning,DBLP:conf/scm/LieCDK89} or a text editor\,\cite{DBLP:journals/ibmrd/Kruskal84}), but before effective and scalable concepts from SPLE research for managing variability have been established. The virtual platform can be seen as a variation control system.

\looseness=-1
The large majority of product-line migration techniques focuses on detecting and analyzing commonalities and variabilities of the cloned variants, together with feature identification and location, as shown in Assuncao et al.'s recent mapping study based on 119 papers\,\cite{Assuncao:2017:fl}. Case studies of manual migration\,\cite{Hetrick2006,Jepsen2007,jacob2020apogamesmigration,Kastner2007,Kolb2006,Stallinger:2011:MTE:1985484.1985490} also exist. These illustrate the difficulties and huge efforts of recovering important information (features and clone relationships) that was never recorded during \co, supporting our approach of recording such information early.
Finally, many works focus on migrating a \emph{single system} into a configurable, product-line platform\,\cite{schulze.ea:2012:refactoring,Kolb2006,Kastner2007,cleland-huang:2020:drones}, typically proposing refactoring techniques. Wille et al.\,\cite{wille2017extractive} use variability mining to generate transformational rules for creating delta-oriented product lines.

Others focus on evolving software platforms.
Liebig et al.\,\cite{liebig.ea:2015:morpheus} present variability-aware sound refactorings (rename identifier, extract function, inline function) for evolving a platform by preserving the variants. Rabiser et al.\,\cite{rabiser2016prototype} present an approach for managing clones at product, component, and feature, and define 5 consistency levels to monitor co-evolving clones. Ignaim et al.\,\cite{ignaim2018systematic} present an extractive approach to engineer cloned variants into systematic reuse.
Neves et al.\,\cite{neves2015safe} propose a set of operators for safe platform evolution.
In contrast to our operationally defined operators, these operators are defined on an abstract level, based on their pre- and post-conditions; implementing them is left to the user. Incorporating safe evolution or Morpheus' refacting in the virtual platform is a valuable future work.

\section{Conclusion}
\looseness=-1
\noindent
We designed, formalized, and prototyped the virtual platform---a framework that exploits a spectrum between the two extremes ad hoc \co and fully integrated
platform, supporting both kinds of development. Based on the number of variants, organizations can decide to use only a subset of all the variability concepts typically required for an integrated platform, fostering flexibility and innovation, starting with \co and incrementally scaling the
development. This realizes incremental benefits for incremental investments and even allows to use \co when a platform is already established, to support a more agile development. Another core novelty is that, instead of trying to expensively recover relevant meta-data (e.g., features, feature locations, and clone traces), the virtual platform fosters recording it early.
For instance, developers typically know the feature they are implementing, but usually do not record it. The virtual platform records such meta-data and exploits it for the transition, providing operators that developers can use to handle variability.
Our evaluation shows that the additional costs are low compared to the benefits.

\looseness=-1
We see several promising directions of future work.
By allowing developers to continuously record feature meta-data, the virtual platform  paves the way for software analyses that rely on this data.
One example is support for the safe evolution of product line platforms\,\cite{neves2015safe}, which could be extended to support systems in our intermediate governance levels.
Specifying  our operators in the framework of software product line transformations\,\cite{taentzer2017transformations,struber2018taming, chechik2016perspectives} would make them amenable to conflict and dependency analysis\,\cite{lambers2018multi}, a versatile formal analysis with applications in the coordination of evolution processes.
Many of the virtual platform's operators (e.g., those related to change propagation) lead to non-trivial changes of the codebase.
To increase developer trust and optimize accuracy, an important challenge is to keep the ``human in the loop'', which we aim to address by exploring dedicated user interfaces.
By integrating the virtual platform with available annotation systems\,\cite{schwarz2020annotations}, we could facilitate inspection of the available feature mappings.
Offering a ``preview mode'' would allow to inspect and interact with the changes arising from a planned operator invocation. \edit{Providing a dedicated operator to integrate cloned features is another future direction.}
Other directions are to support configuration of variants by selecting features, offering views\,\cite{stanciulescu.ea:2016:vts}, and providing visualizations (e.g., dashboards\,\cite{andam2017florida,entekhabi2019featuredashboard}).
Finally, recommender systems that learn from the meta-data and support developers handling features and assets could further encourage using features in software engineering\,\cite{abukwaik2018traceability}.

\parhead{Acknowledgment.} Swedish Research Council (257822902), Vinnova Sweden (2016-02804), and the Wallenberg Academy.

\bibliographystyle{IEEEtran}
\bibliography{main}

\begin{thebibliography}{10}
\providecommand{\url}[1]{#1}
\csname url@samestyle\endcsname
\providecommand{\newblock}{\relax}
\providecommand{\bibinfo}[2]{#2}
\providecommand{\BIBentrySTDinterwordspacing}{\spaceskip=0pt\relax}
\providecommand{\BIBentryALTinterwordstretchfactor}{4}
\providecommand{\BIBentryALTinterwordspacing}{\spaceskip=\fontdimen2\font plus
\BIBentryALTinterwordstretchfactor\fontdimen3\font minus
  \fontdimen4\font\relax}
\providecommand{\BIBforeignlanguage}[2]{{%
\expandafter\ifx\csname l@#1\endcsname\relax
\typeout{** WARNING: IEEEtran.bst: No hyphenation pattern has been}%
\typeout{** loaded for the language `#1'. Using the pattern for}%
\typeout{** the default language instead.}%
\else
\language=\csname l@#1\endcsname
\fi
#2}}
\providecommand{\BIBdecl}{\relax}
\BIBdecl

\bibitem{dubinski.ea:2013:cloning}
Y.~Dubinsky, J.~Rubin, T.~Berger, S.~Duszynski, M.~Becker, and K.~Czarnecki,
  ``An exploratory study of cloning in industrial software product lines,'' in
  \emph{CSMR}, 2013.

\bibitem{stuanciulescu2015forked}
{\c{S}}.~St{\u{a}}nciulescu, S.~Schulze, and A.~W{\k{a}}sowski, ``Forked and
  integrated variants in an open-source firmware project,'' in \emph{ICSME},
  2015.

\bibitem{businge.ea:2018:appfamilies}
J.~Businge, O.~Moses, S.~Nadi, E.~Bainomugisha, and T.~Berger, ``Clone-based
  variability management in the android ecosystem,'' in \emph{ICSME}, 2018.

\bibitem{jacob2020costs}
J.~Krueger and T.~Berger, ``An empirical analysis of the costs of clone- and
  platform-oriented software reuse,'' in \emph{FSE}, 2020.

\bibitem{lodewijksanalysis}
N.~Lodewijks, ``Analysis of a clone-and-own industrial automation system: An
  exploratory study,'' in \emph{SATToSE}, 2017.

\bibitem{berger2013survey}
T.~Berger, R.~Rublack, D.~Nair, J.~M. Atlee, M.~Becker, K.~Czarnecki, and
  A.~Wasowski, ``A survey of variability modeling in industrial practice,'' in
  \emph{VaMoS}, 2013.

\bibitem{clements.ea:01:software}
P.~Clements and L.~Northrop, \emph{Software Product Lines: Practices and
  Patterns}, 2001.

\bibitem{czarnecki.ea:00:generative}
K.~Czarnecki and U.~W. Eisenecker, \emph{Generative Programming: Methods,
  Tools, and Applications}, 2000.

\bibitem{linden.ea:2007:practices}
F.~J. van~der Linden, K.~Schmid, and E.~Rommes, \emph{Software Product Lines in
  Action: The Best Industrial Practice in Product Line Engineering}, 2007.

\bibitem{apel2013software}
S.~Apel, D.~Batory, C.~K{\"a}stner, and G.~Saake, in \emph{Feature-Oriented
  Software Product Lines}.\hskip 1em plus 0.5em minus 0.4em\relax Springer,
  2013.

\bibitem{berger2020state}
T.~Berger, J.-P. Stegh{\"o}fer, T.~Ziadi, J.~Robin, and J.~Martinez, ``The
  state of adoption and the challenges of systematic variability management in
  industry,'' \emph{Empirical Software Engineering}, vol.~25, no.~3, pp.
  1755--1797, 2020.

\bibitem{kang.ea:1990:foda}
K.~Kang, S.~Cohen, J.~Hess, W.~Nowak, and S.~Peterson, ``Feature-oriented
  domain analysis {(FODA)} feasibility study,'' Carnegie-Mellon University,
  Pittsburgh, PA, USA, Tech. Rep., 1990.

\bibitem{damir2019principles}
D.~Nesic, J.~Krueger, S.~Stanciulescu, and T.~Berger, ``Principles of feature
  modeling,'' in \emph{FSE}, 2019.

\bibitem{berger2010featuretocode}
T.~Berger, S.~She, R.~Lotufo, K.~Czarnecki, and A.~Wasowski, ``Feature-to-code
  mapping in two large product lines,'' in \emph{SPLC}, 2010, extended
  Abstract.

\bibitem{linsbauer2013recovering}
L.~Linsbauer, E.~R. Lopez-Herrejon, and A.~Egyed, ``Recovering traceability
  between features and code in product variants,'' in \emph{SPLC}, 2013.

\bibitem{bashroush2017case}
R.~Bashroush, M.~Garba, R.~Rabiser, I.~Groher, and G.~Botterweck, ``Case tool
  support for variability management in software product lines,'' \emph{ACM
  Computing Surveys (CSUR)}, vol.~50, no.~1, pp. 1--45, 2017.

\bibitem{Stallinger:2011:MTE:1985484.1985490}
F.~Stallinger, R.~Neumann, R.~Schossleitner, and S.~Kriener, ``Migrating
  towards evolving software product lines: Challenges of an {SME} in a core
  customer-driven industrial systems engineering context,'' in \emph{PLEASE},
  2011.

\bibitem{Jepsen2007}
H.~P. Jepsen, J.~G. Dall, and D.~Beuche, ``Minimally invasive migration to
  software product lines,'' in \emph{SPLC}, 2007.

\bibitem{assunccao2017reengineering}
W.~K.~G. Assun{\c{c}}{\~a}o, R.~E. Lopez-Herrejon, L.~Linsbauer, S.~R.
  Vergilio, and A.~Egyed, ``Reengineering legacy applications into software
  product lines: a systematic mapping,'' \emph{Empirical Software Engineering},
  vol.~22, no.~6, pp. 2972--3016, 2017.

\bibitem{jacob2020apogamesmigration}
J.~Krueger and T.~Berger, ``Activities and costs of re-engineering cloned
  variants into an integrated platform,'' in \emph{VaMoS}, 2020.

\bibitem{variclouds}
J.~Martinez, T.~Ziadi, T.~F. Bissyand{\'{e}}, J.~Klein, and Y.~L. Traon, ``Name
  suggestions during feature identification: The variclouds approach,'' in
  \emph{SPLC}, 2016.

\bibitem{zhou2018identifying}
S.~Zhou, S.~St\u{a}nciulescu, O.~Le{\ss}enich, Y.~Xiong, A.~Wasowski, and
  C.~K{\"a}stner, ``Identifying features in forks,'' in \emph{ICSE}, 2018.

\bibitem{bennasr17}
S.~B. Nasr, G.~B{\'{e}}can, M.~Acher, J.~B.~F. Filho, N.~Sannier, B.~Baudry,
  and J.~Davril, ``Automated extraction of product comparison matrices from
  informal product descriptions,'' \emph{Journal of Systems and Software}, vol.
  124, pp. 82--103, 2017.

\bibitem{Rubin2013}
J.~Rubin and M.~Chechik, ``{A Survey of Feature Location Techniques},'' in
  \emph{Domain Engineering}, 2013, pp. 29--58.

\bibitem{dit2013feature}
B.~Dit, M.~Revelle, M.~Gethers, and D.~Poshyvanyk, ``Feature location in source
  code: a taxonomy and survey,'' \emph{Journal of software: Evolution and
  Process}, vol.~25, no.~1, pp. 53--95, 2013.

\bibitem{michelon2021hybrid}
G.~K. Michelon, L.~Linsbauer, W.~K. Assun{\c{c}}{\~a}o, S.~Fischer, and
  A.~Egyed, ``A hybrid feature location technique for re-engineering single
  systems into software product lines,'' in \emph{VaMoS}, 2021.

\bibitem{kastner2011variability}
C.~K{\"a}stner, A.~Dreiling, and K.~Ostermann, ``Variability mining with
  leadt,'' \emph{Tec. Rep., Philipps Univ. Marburg}, 2011.

\bibitem{kastner2013variability}
C.~K{\"{a}}stner, A.~Dreiling, and K.~Ostermann, ``Variability mining:
  Consistent semi-automatic detection of product-line features,'' \emph{IEEE
  Transactions on Software Engineering}, vol.~40, no.~1, pp. 67--82, 2013.

\bibitem{roy2007survey}
C.~K. Roy and J.~R. Cordy, ``A survey on software clone detection research,''
  \emph{Queen’s School of Computing TR}, vol. 541, no. 115, pp. 64--68, 2007.

\bibitem{rattan2013software}
D.~Rattan, R.~Bhatia, and M.~Singh, ``Software clone detection: A systematic
  review,'' \emph{Information and Software Technology}, vol.~55, no.~7, pp.
  1165--1199, 2013.

\bibitem{wang.ea:2013:howfeaturelocationdone}
J.~Wang, X.~Peng, Z.~Xing, and W.~Zhao, ``How developers perform feature
  location tasks: a human-centric and process-oriented exploratory study,''
  \emph{Journal of Software: Evolution and Process}, vol.~25, no.~11, pp.
  1193--1224, 2013.

\bibitem{gruner2020incremental}
S.~Gr{\"u}ner, A.~Burger, T.~Kantonen, and J.~R{\"u}ckert, ``Incremental
  migration to software product line engineering,'' in \emph{SPLC}, 2020.

\bibitem{kruger2020promote}
J.~Kr{\"u}ger, W.~Mahmood, and T.~Berger, ``Promote-pl: a round-trip
  engineering process model for adopting and evolving product lines,'' in
  \emph{SPLC}, 2020.

\bibitem{zhang2013variability}
B.~Zhang, M.~Becker, T.~Patzke, K.~Sierszecki, and J.~E. Savolainen,
  ``Variability evolution and erosion in industrial product lines: a case
  study,'' in \emph{SPLC}, 2013.

\bibitem{antkiewicz2014flexible}
M.~Antkiewicz, W.~Ji, T.~Berger, K.~Czarnecki, T.~Schmorleiz, R.~L{\"a}mmel,
  t.~St{\u{a}}nciulescu, A.~Wasowski, and I.~Schaefer, ``Flexible product line
  engineering with a virtual platform,'' in \emph{ICSE-NIER}, 2014.

\bibitem{Fogdal2016}
T.~Fogdal, H.~Scherrebeck, J.~Kuusela, M.~Becker, and B.~Zhang, ``Ten years of
  product line engineering at danfoss: lessons learned and way ahead,'' in
  \emph{SPLC}, 2016.

\bibitem{appendix:impl}
``Virtual platform prototype,''
  \url{https://bitbucket.org/easelab/workspace/projects/VP}.

\bibitem{appendix:online}
``Appendix,'' \url{https://bitbucket.org/easelab/2021-icse-vponlineappendix}.

\bibitem{laguna2013systematic}
M.~A. Laguna and Y.~Crespo, ``A systematic mapping study on software product
  line evolution: From legacy system reengineering to product line
  refactoring,'' \emph{Science of Computer Programming}, vol.~78, no.~8, pp.
  1010--1034, 2013.

\bibitem{passos.ea:2018:tse}
L.~Passos, R.~Queiroz, M.~Mukelabai, T.~Berger, S.~Apel, K.~Czarnecki, and
  J.~Padilla, ``A study of feature scattering in the linux kernel,'' \emph{IEEE
  Transactions on Software Engineering}, vol.~47, pp. 146--164, 2021.

\bibitem{passos.ea:2015:scattering}
L.~Passos, J.~Padilla, T.~Berger, S.~Apel, K.~Czarnecki, and M.~T. Valente,
  ``Feature scattering in the large: A longitudinal study of {L}inux kernel
  device drivers,'' in \emph{MODULARITY}, 2015.

\bibitem{berger.ea:2015:feature}
T.~Berger, D.~Lettner, J.~Rubin, P.~Grünbacher, A.~Silva, M.~Becker,
  M.~Chechik, and K.~Czarnecki, ``What is a feature? a qualitative study of
  features in industrial software product lines,'' in \emph{SPLC}, 2015.

\bibitem{kruger2018towards}
J.~Kr{\"u}ger, W.~Gu, H.~Shen, M.~Mukelabai, R.~Hebig, and T.~Berger, ``Towards
  a better understanding of software features and their characteristics: a case
  study of marlin,'' in \emph{VaMoS}, 2018.

\bibitem{kruger2019my}
J.~Kr{\"u}ger, M.~Mukelabai, W.~Gu, H.~Shen, R.~Hebig, and T.~Berger, ``Where
  is my feature and what is it about? a case study on recovering feature
  facets,'' \emph{Journal of Systems and Software}, vol. 152, pp. 239--253,
  2019.

\bibitem{KruegerNF+17}
J.~Kr{\"u}ger, L.~Nell, W.~Fenske, G.~Saake, and T.~Leich, ``{Finding Lost
  Features in Cloned Systems},'' in \emph{SPLC}, 2017.

\bibitem{berger2013study}
T.~Berger, S.~She, R.~Lotufo, A.~Wasowski, and K.~Czarnecki, ``A study of
  variability models and languages in the systems software domain,'' \emph{IEEE
  Transactions on Software Engineering}, vol.~39, no.~12, pp. 1611--1640, 2013.

\bibitem{sincero.ea:osspl}
J.~Sincero, H.~Schirmeier, W.~Schr{\"o}der-Preikschat, and O.~Spinczyk, ``{Is
  The Linux Kernel a {S}oftware {P}roduct {L}ine?}'' in \emph{{SPLC-OSSPL}},
  2007.

\bibitem{Hetrick2006}
W.~A. Hetrick, C.~W. Krueger, and J.~G. Moore, ``Incremental return on
  incremental investment: Engenio's transition to software product line
  practice,'' in \emph{OOPSLA}, 2006.

\bibitem{bilic.ea:2020:volvo}
D.~Bilic, D.~Sundmark, W.~Afzal, P.~Wallin, A.~Causevic, C.~Amlinger, and
  D.~Barkah, ``Towards a model-driven product line engineering process: An
  industrial case study,'' in \emph{ISEC}, 2020.

\bibitem{rubin.ea:2013:framework}
J.~Rubin, K.~Czarnecki, and M.~Chechik, ``Managing cloned variants: A framework
  and experience,'' in \emph{SPLC}, 2013.

\bibitem{Rubin2015}
------, ``Cloned product variants: from ad-hoc to managed software product
  lines,'' \emph{International Journal on Software Tools for Technology
  Transfer}, vol.~5, no.~17, pp. 627--646, 2015.

\bibitem{fischer.ea:2014:ecco}
S.~Fischer, L.~Linsbauer, R.~E. Lopez{-}Herrejon, and A.~Egyed, ``Enhancing
  clone-and-own with systematic reuse for developing software variants,'' in
  \emph{ICSME}, 2014.

\bibitem{MartinezBut4Reuse}
J.~Martinez, T.~Ziadi, T.~F. Bissyandé, J.~Klein, and Y.~L. Traon, ``Bottom-up
  technologies for reuse: Automated extractive adoption of software product
  lines,'' in \emph{ICSE-C}, 2017.

\bibitem{variantsync:2016}
T.~Pfofe, T.~Thüm, S.~Schulze, W.~Fenske, and I.~Schaefer, ``Synchronizing
  software variants with variantsync,'' in \emph{SPLC}, 2016.

\bibitem{montalvillo2015tuning}
L.~Montalvillo and O.~D{\'\i}az, ``Tuning github for spl development: branching
  models \& repository operations for product engineers,'' in \emph{SPLC},
  2015.

\bibitem{apel.ea:2009:featurehouse}
S.~Apel, C.~K{\"{a}}stner, and C.~Lengauer, ``Featurehouse:
  Language-independent, automated software composition,'' in \emph{ICSE}, 2009.

\bibitem{she2010variability}
S.~She, R.~Lotufo, T.~Berger, A.~Wasowski, and K.~Czarnecki, ``The variability
  model of the linux kernel.'' \emph{VaMoS}, 2010.

\bibitem{ji2015maintaining}
W.~Ji, T.~Berger, M.~Antkiewicz, and K.~Czarnecki, ``Maintaining feature
  traceability with embedded annotations,'' in \emph{SPLC}, 2015.

\bibitem{struber2020variability}
D.~Str{\"u}ber, A.~Anjorin, and T.~Berger, ``Variability representations in
  class models: An empirical assessment,'' in \emph{MODELS}, 2020.

\bibitem{bkak2016clafer}
K.~B{\k{a}}k, Z.~Diskin, M.~Antkiewicz, K.~Czarnecki, and A.~W{\k{a}}sowski,
  ``Clafer: unifying class and feature modeling,'' \emph{Software \& Systems
  Modeling}, vol.~15, no.~3, pp. 811--845, 2016.

\bibitem{antkiewicz2013clafer}
M.~Antkiewicz, K.~Bak, A.~Murashkin, R.~Olaechea, J.~Hui, and K.~Czarnecki,
  ``Clafer tools for product line engineering.'' in \emph{SPLC Workshops},
  2013.

\bibitem{koschke2007survey}
R.~Koschke, ``Survey of research on software clones,'' in \emph{Dagstuhl
  Seminar Proceedings}.\hskip 1em plus 0.5em minus 0.4em\relax Schloss
  Dagstuhl-Leibniz-Zentrum f{\"u}r Informatik, 2007.

\bibitem{sajnani2016sourcerercc}
H.~Sajnani, V.~Saini, J.~Svajlenko, C.~K. Roy, and C.~V. Lopes, ``Sourcerercc:
  Scaling code clone detection to big-code,'' in \emph{ICSE}, 2016.

\bibitem{martinez2017espla}
J.~Martinez, W.~K.~G. Assun{\c{c}}{\~a}o, and T.~Ziadi, ``Espla: A catalog of
  extractive spl adoption case studies,'' in \emph{SPLC}, 2017.

\bibitem{struber2019facing}
D.~Str{\"u}ber, M.~Mukelabai, J.~Kr{\"u}ger, S.~Fischer, L.~Linsbauer,
  J.~Martinez, and T.~Berger, ``Facing the truth: benchmarking the techniques
  for the evolution of variant-rich systems,'' in \emph{SPLC}, 2019.

\bibitem{berger2019dagstuhl}
T.~Berger, M.~Chechik, T.~Kehrer, and M.~Wimmer, ``Software evolution in time
  and space: Unifying version and variability management (dagstuhl seminar
  19191),'' in \emph{Dagstuhl Reports}.\hskip 1em plus 0.5em minus 0.4em\relax
  Schloss Dagstuhl -- Leibniz-Zentrum fuer Informatik, 2019.

\bibitem{catal2009systematic}
C.~Catal and B.~Diri, ``A systematic review of software fault prediction
  studies,'' \emph{Expert systems with applications}, vol.~36, no.~4, pp.
  7346--7354, 2009.

\bibitem{struder2020feature}
S.~Str{\"u}der, M.~Mukelabai, D.~Str{\"u}ber, and T.~Berger, ``Feature-oriented
  defect prediction,'' in \emph{SPLC}, 2020.

\bibitem{Linsbauer2017}
L.~Linsbauer, T.~Berger, and P.~Gr\"{u}nbacher, ``A classification of variation
  control systems,'' in \emph{GPCE}, 2017.

\bibitem{linsbauer2020varcs}
L.~Linsbauer, F.~Schwaegerl, T.~Berger, and P.~Gruenbacher, ``Concepts of
  variation control systems,'' \emph{Journal of Systems and Software}, vol.
  171, p. 110796, 2021.

\bibitem{munch.ea:1993:uniformversioning}
B.~P. Munch, J.-O. Larsen, B.~Gulla, R.~Conradi, and E.-A. Karlsson, ``Uniform
  versioning: The change-oriented model,'' in \emph{SCM}, 1993.

\bibitem{DBLP:conf/scm/LieCDK89}
A.~Lie, R.~Conradi, T.~Didriksen, and E.~Karlsson, ``Change oriented versioning
  in a software engineering database,'' in \emph{SCM}, 1989, pp. 56--65.

\bibitem{DBLP:journals/ibmrd/Kruskal84}
V.~J. Kruskal, ``Managing multi-version programs with an editor,'' \emph{{IBM}
  Journal of Research and Development}, vol.~28, no.~1, pp. 74--81, 1984.

\bibitem{Assuncao:2017:fl}
W.~K.~G. Assun{\c{c}}{\~{a}}o, R.~E. Lopez{-}Herrejon, L.~Linsbauer, S.~R.
  Vergilio, and A.~Egyed, ``Reengineering legacy applications into software
  product lines: A systematic mapping,'' \emph{Empirical Software Engineering},
  vol.~22, no.~6, pp. 2972--3016, 2017.

\bibitem{Kastner2007}
C.~K{\"{a}}stner, S.~Apel, and D.~S. Batory, ``A case study implementing
  features using aspectj,'' in \emph{SPLC}, 2007.

\bibitem{Kolb2006}
R.~Kolb, D.~Muthig, T.~Patzke, and K.~Yamauchi, ``Refactoring a legacy
  component for reuse in a software product line: a case study,'' \emph{Journal
  of Software Maintenance}, vol.~18, no.~2, pp. 109--132, 2006.

\bibitem{schulze.ea:2012:refactoring}
S.~Schulze, T.~Th\"{u}m, M.~Kuhlemann, and G.~Saake, ``Variant-preserving
  refactoring in feature-oriented software product lines,'' in \emph{VaMoS},
  2012.

\bibitem{cleland-huang:2020:drones}
J.~Cleland-Huang, A.~Agrawal, M.~N.~A. Islam, E.~Tsai, M.~Van~Speybroeck, and
  M.~Vierhauser, ``Requirements-driven configuration of emergency response
  missions with small aerial vehicles,'' in \emph{SPLC}, 2020.

\bibitem{wille2017extractive}
D.~Wille, T.~Runge, C.~Seidl, and S.~Schulze, ``Extractive software product
  line engineering using model-based delta module generation,'' in
  \emph{VaMoS}, 2017.

\bibitem{liebig.ea:2015:morpheus}
J.~Liebig, A.~Janker, F.~Garbe, S.~Apel, and C.~Lengauer, ``Morpheus:
  Variability-aware refactoring in the wild,'' in \emph{ICSE}, 2015.

\bibitem{rabiser2016prototype}
D.~Rabiser, P.~Gr{\"u}nbacher, H.~Pr{\"a}hofer, and F.~Angerer, ``A
  prototype-based approach for managing clones in clone-and-own product
  lines,'' in \emph{Proceedings of the 20th International Systems and Software
  Product Line Conference}, 2016, pp. 35--44.

\bibitem{ignaim2018systematic}
K.~Ignaim, J.~M. Fernandes, A.~L. Ferreira, and J.~Seidel, ``A systematic
  reuse-based approach for customized cloned variants,'' in \emph{QUATIC},
  2018.

\bibitem{neves2015safe}
L.~Neves, P.~Borba, V.~Alves, L.~Turnes, L.~Teixeira, D.~Sena, and U.~Kulesza,
  ``Safe evolution templates for software product lines,'' \emph{Journal of
  Systems and Software}, vol. 106, pp. 42--58, 2015.

\bibitem{taentzer2017transformations}
G.~Taentzer, R.~Salay, D.~Str{\"u}ber, and M.~Chechik, ``Transformations of
  software product lines: A generalizing framework based on category theory,''
  in \emph{MODELS}, 2017.

\bibitem{struber2018taming}
D.~Str{\"u}ber, S.~Peldszus, and J.~J{\"u}rjens, ``Taming multi-variability of
  software product line transformations,'' in \emph{FASE}, 2018.

\bibitem{chechik2016perspectives}
M.~Chechik, M.~Famelis, R.~Salay, and D.~Str{\"u}ber, ``Perspectives of model
  transformation reuse,'' in \emph{IFM}, 2016.

\bibitem{lambers2018multi}
L.~Lambers, D.~Str{\"u}ber, G.~Taentzer, K.~Born, and J.~Huebert,
  ``Multi-granular conflict and dependency analysis in software engineering
  based on graph transformation,'' in \emph{ICSE}, 2018.

\bibitem{schwarz2020annotations}
T.~Schwarz, W.~Mahmood, and T.~Berger, ``A common notation and tool support for
  embedded feature annotations,'' in \emph{SPLC}, 2020.

\bibitem{stanciulescu.ea:2016:vts}
S.~Stanciulescu, T.~Berger, E.~Walkingshaw, and A.~W\k{a}sowski, ``Concepts,
  operations, and feasibility of a projection-based variation control system,''
  in \emph{ICSME}, 2016.

\bibitem{andam2017florida}
B.~Andam, A.~Burger, T.~Berger, and M.~R.~V. Chaudron, ``Florida: Feature
  location dashboard for extracting and visualizing feature traces,'' in
  \emph{VaMoS}, 2017.

\bibitem{entekhabi2019featuredashboard}
S.~Entekhabi, A.~Solback, J.-P. Stegh\"ofer, and T.~Berger, ``Visualization of
  feature locations with the tool featuredashboard,'' in \emph{SPLC, Tools
  Track}, 2019.

\bibitem{abukwaik2018traceability}
H.~Abukwaik, A.~Burger, B.~Andam, and T.~Berger, ``Semi-automated feature
  traceability with embedded annotations,'' in \emph{ICSME}, 2018.

\end{thebibliography}

\end{document}